\documentclass[twoside,twocolumn,9pt]{article}
\usepackage{extsizes}
\usepackage[super,sort&compress,comma]{natbib}
\usepackage[version=3]{mhchem}
\usepackage[left=1.5cm, right=1.5cm, top=1.785cm, bottom=2.0cm]{geometry}
\usepackage{balance}
\usepackage{times,mathptmx}
\usepackage{sectsty}
\usepackage{graphicx}
\usepackage{lastpage}
\usepackage[format=plain,justification=justified,singlelinecheck=false,font={stretch=1.125,small,sf},labelfont=bf,labelsep=space]{caption}
\usepackage{float}
\usepackage{fancyhdr}
\usepackage{fnpos}
\usepackage[english]{babel}
\usepackage{array}
\usepackage{droidsans}
\usepackage{charter}
\usepackage[T1]{fontenc}
\usepackage[usenames,dvipsnames]{xcolor}
\usepackage{setspace}
\usepackage[compact]{titlesec}
\usepackage{hyperref}

\usepackage{epstopdf}

\definecolor{cream}{RGB}{222,217,201}

\begin{document}

\pagestyle{fancy}
\thispagestyle{plain}
\fancypagestyle{plain}{

\fancyhead[C]{\includegraphics[width=18.5cm]{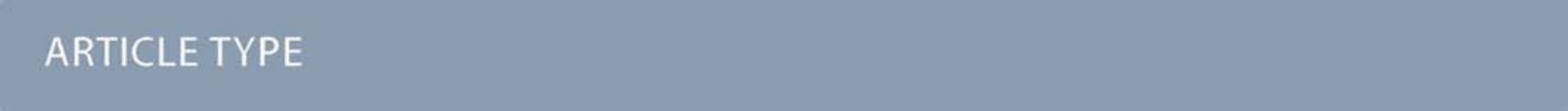}}
\fancyhead[L]{\hspace{0cm}\vspace{1.5cm}\includegraphics[height=30pt]{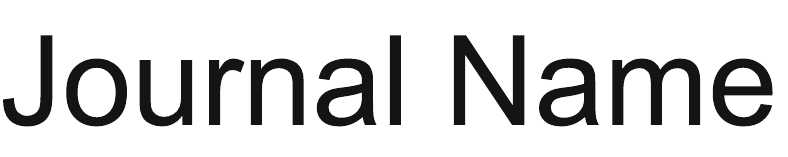}}
\fancyhead[R]{\hspace{0cm}\vspace{1.7cm}\includegraphics[height=55pt]{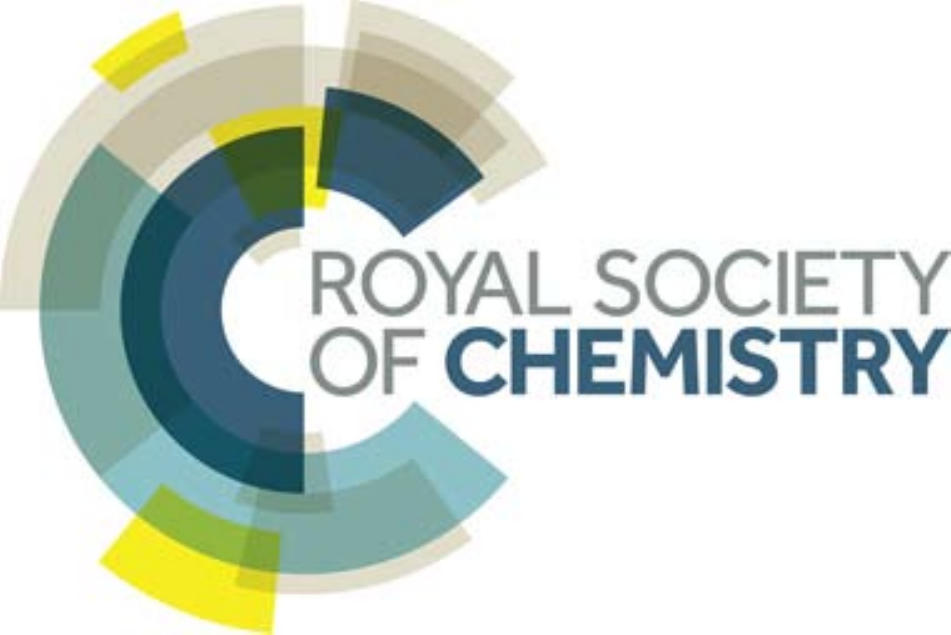}}
\renewcommand{\headrulewidth}{0pt}
}

\makeFNbottom
\makeatletter
\renewcommand\LARGE{\@setfontsize\LARGE{15pt}{17}}
\renewcommand\Large{\@setfontsize\Large{12pt}{14}}
\renewcommand\large{\@setfontsize\large{10pt}{12}}
\renewcommand\footnotesize{\@setfontsize\footnotesize{7pt}{10}}
\makeatother

\renewcommand{\thefootnote}{\fnsymbol{footnote}}
\renewcommand\footnoterule{\vspace*{1pt}%
\color{cream}\hrule width 3.5in height 0.4pt \color{black}\vspace*{5pt}}
\setcounter{secnumdepth}{5}

\makeatletter
\renewcommand\@biblabel[1]{#1}
\renewcommand\@makefntext[1]%
{\noindent\makebox[0pt][r]{\@thefnmark\,}#1}
\makeatother
\renewcommand{\figurename}{\small{Fig.}~}
\sectionfont{\sffamily\Large}
\subsectionfont{\normalsize}
\subsubsectionfont{\bf}
\setstretch{1.125} 
\setlength{\skip\footins}{0.8cm}
\setlength{\footnotesep}{0.25cm}
\setlength{\jot}{10pt}
\titlespacing*{\section}{0pt}{4pt}{4pt}
\titlespacing*{\subsection}{0pt}{15pt}{1pt}

\fancyfoot{}
\fancyfoot[LO,RE]{\vspace{-7.1pt}\includegraphics[height=9pt]{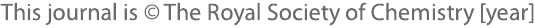}}
\fancyfoot[CO]{\vspace{-7.1pt}\hspace{13.2cm}\includegraphics{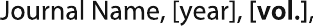}}
\fancyfoot[CE]{\vspace{-7.2pt}\hspace{-14.2cm}\includegraphics{head_foot/RF}}
\fancyfoot[RO]{\footnotesize{\sffamily{1--\pageref{LastPage} ~\textbar  \hspace{2pt}\thepage}}}
\fancyfoot[LE]{\footnotesize{\sffamily{\thepage~\textbar\hspace{3.45cm} 1--\pageref{LastPage}}}}
\fancyhead{}
\renewcommand{\headrulewidth}{0pt}
\renewcommand{\footrulewidth}{0pt}
\setlength{\arrayrulewidth}{1pt}
\setlength{\columnsep}{6.5mm}
\setlength\bibsep{1pt}

\makeatletter
\newlength{\figrulesep}
\setlength{\figrulesep}{0.5\textfloatsep}

\newcommand{\topfigrule}{\vspace*{-1pt}%
\noindent{\color{cream}\rule[-\figrulesep]{\columnwidth}{1.5pt}} }

\newcommand{\botfigrule}{\vspace*{-2pt}%
\noindent{\color{cream}\rule[\figrulesep]{\columnwidth}{1.5pt}} }

\newcommand{\dblfigrule}{\vspace*{-1pt}%
\noindent{\color{cream}\rule[-\figrulesep]{\textwidth}{1.5pt}} }

\makeatother

\twocolumn[
  \begin{@twocolumnfalse}
\vspace{3cm}
\sffamily
\begin{tabular}{m{4.5cm} p{13.5cm} }

\includegraphics{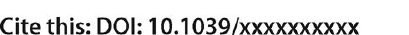} & \noindent\LARGE{\textbf{Entropy production in thermal phase separation: a kinetic-theory approach }} \\
\vspace{0.3cm} & \vspace{0.3cm} \\

 & \noindent\large{Yudong Zhang,\textit{$^{ab}$} Aiguo Xu,$^{\ast}$\textit{$^{acd}$} Guangcai Zhang,\textit{$^a$} Yanbiao Gan,\textit{$^{ef}$}, Zhihua Chen, \textit{$^b$} and Sauro Succi \textit{$^{gh}$} } \\

\includegraphics{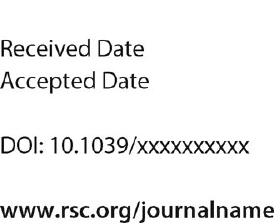} & \noindent\normalsize{Entropy production during the process of thermal phase-separation of multiphase flows is investigated by means of a discrete Boltzmann kinetic model. The entropy production rate is found to increase during the spinodal decomposition stage and to decrease during the domain growth stage,
attaining its maximum at the crossover between the two. Such behaviour provides a natural criterion to identify and discriminate between the two regimes. Furthermore, the effects of heat conductivity, viscosity and surface tension on the entropy production rate are investigated by systematically probing the interplay between non-equilibrium energy and momentum fluxes.
It is found that the entropy production rate due to energy fluxes is
an increasing function of the Prandtl number, while the momentum fluxes
exhibit an opposite trend. On the other hand, both contributions show an increasing trend with surface tension. The present analysis inscribes within the general framework of non-equilibrium thermodynamics and consequently it is expected to be relevant to a broad class of soft-flowing systems far from mechanical and thermal equilibrium.} \\

\end{tabular}

 \end{@twocolumnfalse} \vspace{0.6cm}

  ]

\renewcommand*\rmdefault{bch}\normalfont\upshape
\rmfamily
\section*{}
\vspace{-1cm}


\footnotetext{\textit{$^{a}$National Key Laboratory of Computational Physics, Institute of Applied Physics
and Computational Mathematics, P. O. Box 8009-26, Beijing, China.
E-mail: Xu\_Aiguo@iapcm.ac.cn}}
\footnotetext{\textit{$^{b}$Key Laboratory of Transient Physics, Nanjing University of Science and Technology, Nanjing 210094, China }}
\footnotetext{\textit{$^{c}$State Key Laboratory of Explosion Science and Technology, Beijing Institute of Technology, Beijing 100081, China}}

\footnotetext{\textit{$^{d}$Center for Applied Physics and Technology, MOE Key Center for High Energy Density
Physics Simulations, College of Engineering, Peking University, Beijing 100871, China}}
\footnotetext{\textit{$^{e}$North China Institute of Aerospace Engineering, Langfang 065000, China}}
\footnotetext{\textit{$^{f}$College of Mathematics and Informatics $\&$ FJKLMAA, Fujian Normal
University, Fuzhou 350007, China}}
\footnotetext{\textit{$^{g}$Center for Life Nano Science at La Sapienza, Fondazione Istituto Italiano di Tecnologia, Viale Regina Margherita 295, 00161 Roma, Italy}}
\footnotetext{\textit{$^{h}$Physics Department and Institute for Applied Computational Science, John A. Paulson School of Engineering and Applied Sciences, Harvard University, Oxford Street 29, Cambridge, Massachusetts 02138, USA}}




\section{Introduction}
Phase-separation processes are ubiquitous and crucial to the dynamics of complex flows, such as  polymers melts, colloids,
surfactants, soft glasse, biological materials, to name but a few \cite{Sperling2006,Asadi2011Resistive}.

Understanding the characteristics of the phase separation, so as to control its morphological evolution
is very important for the design of various materials with novel rheological, mechanical, optical, and electrical properties.
For this reason, significant efforts have been devoted to the investigation of the phase separation process, by means of
experimental, theoretical, and numerical methods \cite{Iwashita2006Self,Ye2013Polymer,Yeganeh2014Anomalous}.

Besides their wide range of applications, multiphase flows still raise a major
challenge to fundamental science, notably non-equilibrium thermodynamics, mostly
on account of the major complexity of their interfacial dynamics.
Non-equilibrium thermodynamics is a notoriously difficult subject, especially far from the linear regime where fluxes no longer scale
in linear proportion with the gradients that drive them.
Under such circumstances, analytical solutions are preciously rare and resort to numerical methods becomes imperative.

Many numerical methods have been developed in the past to address these problems, including phase-field, Lagrangian and Eulerian versions of non-ideal Navier-Stokes and others \cite{Xu2015Progress,Succi2018The}.
Despite their broad variety, most of these methods are based on the discretisation of
the macroscopic equations of non-ideal thermo-hydrodynamics with suitable interface boundary conditions.
Although such methods have achieved major progress, they still face with a number of problems whenever interfacial dynamics presents large, localised gradients across complex topologies. In this respect, kinetic theory, being capable, at least in principle, of handling arbitrarily large gradients (large Knudsen numbers), is expected to offer a broader angle attack.
Unfortunately, the cornerstone of kinetic theory, namely the Boltzmann equation, besides being computationally very demanding, does
not easily extend to the dense fluid regime which is relevant to most multiphase flows \cite{Chapman1953The,Succi2018The}.

However, in the last decade, suitable model Boltzmann equations, living in discrete phase-space, have proven capable of incorporating the basic features which control the physics of multiphase flows, namely a non-ideal equation of state, surface tension and disjoining pressure \cite{Gan2015Discrete,Benzi1992The,Succi2018The,Gonnella2007Lattice,Li2016Lattice,Gunstensen1991Lattice,Shan1993Lattice,Shan1994Simulation,Sbragaglia2007Generalized,Falcucci2010Lattice,Swift1995Lattice,Swift1996Lattice,Xu2003Phase,He1999A,Karlin1,Karlin2}.

Some of the models or improved versions thereof, have been successfully used in the simulation of complex fluids such as polymers \cite{Ledesmaaguilar2012Length}, soft glassy materials \cite{Sbragaglia2012The}, liquid crystals \cite{Mackay2013Deformable,Cates2009Lattice}, and porous materials \cite{Vanson2015Unexpected,Vanson2016Transport}.
With the help of those LB multiphase models, a wide variety of multiphase problems, including wetting \cite{Vrancken2012Anisotropic,Li2014Contact,Tang2015Study}, droplet dynamic and evaporation \cite{Li2015Lattice,Cheng2017A,Ledesmaaguilar2014Lattice}, phase transition \cite{Falcucci2010Lattice,Liu2015Double,Luo2015Lattice,Gan2015Discrete}, hydrodynamic instability \cite{He1999A}, etc., have been successfully simulated and investigated.

This has opened up a new computational route to the exploration of multiphase flows, which is precisely the framework this paper inscribes to. More specifically, owing to the detailed information on the equilibrium and non-equilibrium kinetic moments of the discrete Boltzmann distribution, we provide a detailed analysis of the entropy evolution during the process of phase separation, as a function of the main transport coefficients, namely momentum and heat diffusivity, as well as surface tension. It is hoped that this kind of analysis may prove useful to gain further insights into the physics of multiphase flows, as well as of other soft flowing systems, such as gels, foams and emulsions.

As a general study, we are not focusing on a specific fluid, which means that all the parameters in this work are dimensionless.

However, physical units can readily be recovered via the similarity principle.
The remainder of this paper is organised as follows.
Section 2 introduces the DBM for multiphase flows, presents the conversion to dimensionless units, and derives
the expression of entropy production rate.
The liquid-vapor coexistence curves and Laplace law are verified to test the new model.
Section 3 demonstrates the characteristics of entropy production for isothermal and thermal phase separations, as well as
the effects of heat conduction, viscous, and surface tension.
The cooperation and competition between the two main mechanisms, NOEF (Non-Organized-Energy-Flux)
versus NOMF (Non-Organized-Momentum-Flux), for entropy production rate are discussed.
Section 4 concludes the present paper.

\section{Methods and validation}

\subsection{Discrete Boltzmann model for non-ideal fluid}

The prime property of non-ideal fluids is their equation of state (EOS), whose
choice is consequently very important for the DBM multiphase formulation as well.

There is a vast choice of EOS for nonideal fluids \cite{Yuan2006Equations,Kupershtokh2009On}, such as the van der Waals (vdW) \cite{PhaseTrans-book},  Meshalkin-Kaplun (M-K) \cite{Kaplun2003Thermodynamic}, Peng-Robinson (P-R) \cite{Peng1976A}, Redlich-Kwong (R-K) \cite{Redlich1949On}, and Carnahan-Starling (C-S) EOS \cite{Carnahan1969Equation}. Among these, the vdW
EOS is the simplest one and widely used in modeling multiphase flows and other soft flowing systems. In fact, the vdW EOS is able to represent almost all basic types of binary phase diagrams for mixtures only if the size difference between the components is not too large \cite{Harismiadis1994Application}. In this work, as a preliminary study, the vdW theory will be used to describe the EOS of nonideal fluid.

\begin{figure}[h]
\centering
  \includegraphics[height=6cm]{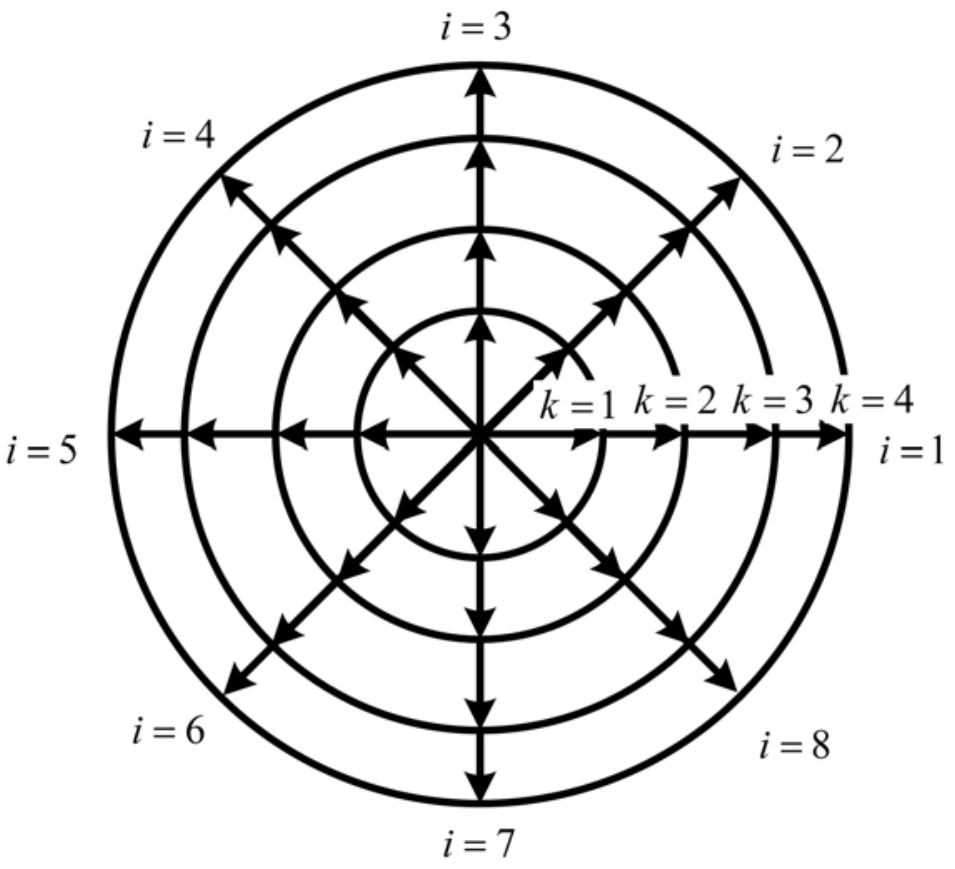}
  \caption{Schematic diagram of the discrete velocity model.}
  \label{fig0-DVM}
\end{figure}

To describe the nonideal EOS and surface tension effects, the collision term in the BGK-Boltzmann equation is augmented with an extra source term \cite{Gonnella2007Lattice}, leading to the following discrete kinetic equation:
\begin{equation}\label{Eq-DBM1}
\frac{{\partial f_{ki}}}{{\partial t}} + {\bf{v}}_{ki} \cdot {\pmb{\nabla}} f_{ki} =  - \frac{1}{\tau }\left( {f_{ki} - {f_{ki}^{eq}}} \right) + I_{ki},
\end{equation}
In the above $f_{ki}$ is distribution function of the discrete velocity ${\bf{v}}_{ki}$  \cite{Watari2003Two} and
$f_{ki}^{eq}$ is the discrete local equilibrium distribution function, where the index $k$ runs over the various energy shells, while $i$ runs over the discrete velocities within each cell (Fig. \ref{fig0-DVM}).

The extra term $I_{ki}$ is used to describe interparticle forces and it is similar to the one introduced
by Klimontovich for nonideal gases \cite{Klimontovich-book}, namely:

\begin{equation}\label{Eq-Iki}
I_{ki} =  - \left[ {A + {\mathbf{B}} \cdot {\mathbf{c}_{ki} } + (C + C_q){c_{ki}^2}} \right]f_{ki}^{eq},
\end{equation}
where $A$, $\mathbf{B}$, $C$, and $C_q$ are four lagrange multipliers encoding the dependence
on macroscopic quantities and their gradients, ${\mathbf{c}}_{ki}={\mathbf{v}}_{ki}-{\mathbf{u}}$ is the peculiar velocity
and $\mathbf{u}$ is macroscopic velocity.

The discretization of particle velocity space and the explicit expression of $f^{eq}_{ki}$
and parameters used in Eq. \eqref{Eq-Iki} are given in previous publications \cite{Gonnella2007Lattice,Gan2012FFT,Gan2015Discrete}.

The DBM used in this work in combination with a vdW EOS is a
kinetic mean-field description of the fluid, lying at an intermediate level between continuum and atomistic dynamics.
As usual, the goal is to combine the best of the two descriptions, namely including the essential microphysics of phase
transitions, without being trapped by unnecessary molecular details.

Besides being consistent with the Onuki model for van der Waals fluid\cite{Onuki2005Dynamic,Onuki2007Dynamic},
\begin{equation}\label{Eq-NS1}
\frac{{\partial \rho }}{{\partial t}} + \nabla  \cdot \left( {\rho {\bf{u}}} \right) = 0,
\end{equation}

\begin{equation}\label{Eq-NS2}
\frac{{\partial \left( {\rho {\bf{u}}} \right)}}{{\partial t}} + \nabla  \cdot \left( {\rho {\bf{uu}} + {P}{\bf{I}}} \right) + \nabla  \cdot \left( {\pmb{{\Lambda}} - {\pmb{\Pi }}} \right) = 0,
\end{equation}

\begin{equation}\label{Eq-NS3}
\frac{{\partial E}}{{\partial t}} + \nabla  \cdot \left( {E{\bf{u}} + {P}{\bf{u}}} \right) + \nabla  \cdot \left[ {({\pmb{\Lambda}} - {\pmb{\Pi}}) \cdot \bf{u} - {{\bf{j}}_q}} \right] = 0,
\end{equation}
in the hydrodynamic limit, the DBM presents more information on the thermodynamic nonequilibrium state and its evolution. The capability of DBM to describe non-equilibrium flows, beyond the Navier-Stokes representation, permits to the study on entropy production in this work.
Here $P = \frac{{\rho T}}{{1 - b \rho}} - a {\rho}^2$ is the van der Waals pressure, ${\bf{I}}$ is the unit tensor, $\pmb{\Lambda}  =  - \left[ {K\rho {\nabla ^2}\rho  + \frac{K}{2}{{\left| {\nabla \rho } \right|}^2} } \right]{\bf{I}} + K\nabla \rho \nabla \rho$ is the contribution of surface tension to the pressure tensor, and $K$ is the coefficient of surface tension. ${\pmb{\Pi }}$  and ${\bf{j}}_q$ are viscous stress and heat flux, respectively, $E=\rho e - a{\rho ^2} + \frac{K}{2}{\left| {\nabla \rho} \right|^2}+\frac{1}{2} \rho u^2$, and $e= DT/2$ is internal energy density without contribution of density gradient where $D$ indicates the spatial dimension.

\subsection{Dimensionless versus physical units}

All the parameters used in the simulation are dimensionless, hence in the following
we illustrate  how to recover the actual physical quantities from the numerical results.

The reference variables are chosen as $\rho_{c}$, $T_{c}$, and $L_{\infty}$, where $\rho_{c}$ and $T_{c}$ are
the critical density and temperature, calculated from the van der Waals equation of state (vdW EOS), respectively. $L_{\infty}$ indicates the global size of the domain.

The relationship between the parameters in DBM and the physical ones are as follows:
\begin{equation}\label{Eq-Dimenless1}
\hat \rho  = \frac{\rho }{{{\rho _c }}}, \hat T = \frac{T}{{{T_c }}}, \hat r_{\alpha} = \frac{r_{\alpha}}{{{L_\infty }}},
\end{equation}

\begin{equation}\label{Eq-Dimenless2}
\left( {\hat t,\hat \tau } \right) = \frac{{\left( {t,\tau } \right)}} {{{L_\infty }/\sqrt {R{T_c}} }}, u = \frac{u}{{\sqrt {R{T_c}} }}, {\hat c_p} = \frac{{{c_p}}}{R},
\end{equation}

\begin{equation}\label{Eq-Dimenless3}
 \hat \mu  = \frac{\mu }{{{\rho _c}{L_\infty }\sqrt {R{T_c}} }}, \hat \kappa  = \frac{\kappa }{{{\rho _c}{L_\infty }R\sqrt {R{T_c}} }}, \hat \sigma  = \frac{\sigma }{{{\rho _c}R{T_c}{L_\infty }}} ,
\end{equation}
where $r_{\alpha}$ indicates the space coordinate in the $\alpha$ direction, $c_p$ is the specific heat at constant
pressure, $\mu$ and $\kappa$ are the viscosity and heat conductivity coefficients, respectively,
$\sigma$ is the surface tension, computed as $\sigma  = K\int_{ - \infty }^\infty  {{{(\frac{{\partial \rho }}{{\partial {r_\alpha }}})}^2}} d{r_\alpha }$.
 Dimensionless variables are denoted by an upper  ``$\wedge$''

The non-dimensional numbers, Prandtl number $Pr$ and capillary number Ca, are defined as
\begin{equation}\label{Eq-Prandtl}
Pr  = \frac{{{c_p}\mu }}{\kappa } = \frac{{{{\hat c}_p}\hat \mu }}{{\hat \kappa }},
\end{equation}
\begin{equation}\label{Eq-Capillary}
Ca = \frac{{u\mu }}{\sigma }{\rm{ = }}\frac{{\hat u\hat \mu }}{{\hat \sigma }}.
\end{equation}

The parameters in EOS and coefficient of surface tension can be determined by the properties of the fluids.
In the latter part of this paper, all quantities are dimensionless, but the symbol ``$\wedge$'' will be dropped for simplicity.

\subsection{Thermodynamic non-equilibrium measurement and entropy production}
The DBM is widely used to study complex flows with significant non-equilibrium effects. It presents two sets as a measure of thermal non-equilibrium (TNE).

 The first goes with the departure between the kinetic
 moments of $f_{ki}$ and the corresponding ones from the equilibrium distribution $f^{eq}_{ki}$, which reads \cite{Xu2012Lattice}
\begin{equation}\label{Eq-Deltaxing-m}
{\pmb{\Delta }}_m^* = {\bf{M}}_m^*\left( {{f_{ki}}} \right) - {\bf{M}}_m^*\left( {f_{ki}^{eq}} \right),
\end{equation}
where ${\bf{M}}_m^*\left( f_{ki} \right)$ indicates the $m$th order kinetic center moment,
\begin{equation}\label{Eq-Mxing-m}
{\bf{M}}_m^*\left( f_{ki} \right){\rm{ = }}\sum\limits_{ki} {{f_{ki}}\underbrace {{({\bf{v}}_{ki}-{\bf{u}})}{({\bf{v}}_{ki}-{\bf{u}})} \cdots {({\bf{v}}_{ki}}-{\bf{u}})}_m}.
\end{equation}
The second set includes the viscous stress and heat flux.

In fact, it can be assumed that the latter set of TNE indicator is contained in the former one. Because the viscous stress and heat flux correspond to ${\pmb{\Delta }}_2^*$ and ${\pmb{\Delta }}_{3,1}^*$, respectively,
where the subscript ``2'' indicates the second-order tensor and ``3,1'' represents the first-order tensor contracted from a third-order tensor.

Taking the zeroth order, first order, and ``$2,0$''th order velocity moments of Eq. (\ref{Eq-DBM1}), delivers
a set of generalized hydrodynamic equations
\begin{equation}\label{Eq-Gns1}
\frac{{\partial \rho }}{{\partial t}} + \nabla  \cdot \left( {\rho {\bf{u}}} \right) = 0,
\end{equation}
\begin{equation}\label{Eq-Gns2}
\frac{{\partial \left( {\rho {\bf{u}}} \right)}}{{\partial t}} + \nabla  \cdot \left( {\rho {\bf{uu}} + {P}{\bf{I}}} \right) + \nabla  \cdot \left( {{\pmb{\Delta }}_{2}^* + {\pmb{\Lambda }}} \right) = 0,
\end{equation}
\begin{equation}\label{Eq-Gns3}
\frac{{\partial E}}{{\partial t}} + \nabla  \cdot \left( {E {\bf{u}} + {P}{\bf{u}}} \right) + \nabla  \cdot \left[ {\left( {{\pmb{\Delta }}_2^* + {\pmb{\Lambda }}} \right) \cdot {\bf{u}} + {\pmb{\Delta }}_{3,1}^*} \right] = 0,
\end{equation}
This is similar to Eqs. (\ref{Eq-NS1})-(\ref{Eq-NS3}), but the viscous stress and heat flux terms are replaced by ${\pmb{\Delta }}_2^*$ and ${\pmb{\Delta }}_{3,1}^*$, respectively. In a previous work, we referred ${\pmb{\Delta }}_2^*$ and ${\pmb{\Delta }}_{3,1}^*$ as to non-organised momentum fluxes (NOMF) and non-organised energy fluxes (NOEF), respectively, and obtained a new entropy equilibrium equation
for single-phase flows with chemical reactions \cite{Zhang2016Kinetic}.

For multiphase flows, the definition of entropy, including the gradient contributions, reads as follows \cite{Onuki2005Dynamic,Onuki2007Dynamic},
 \begin{equation}\label{Eq-Entropy-Sb}
{S_b} = \int {\left( {n s - \frac{1}{2}C{{\left| {\nabla n} \right|}^2}} \right)} d{\bf{r}},
\end{equation}
where the space integrals extend to the entire computational domain, $n$ is the particle number density,
$C$ is a constant and the gradient term represents a decrease of entropy because of inhomogeneity of $n$.

The entropy per particle $s$, can be derived directly from the partition function \cite{PhaseTrans-book}, and reads as follows:
\begin{equation}\label{Eq-Entropy-s-perparticle}
s = k_B{\left[ {\frac{{\partial \left( { T\ln {Z_n}} \right)}}{{\partial T}}} \right]_{VN}}/N,
\end{equation}
where $k_B$ is the Boltzmann constant, $Z_n$ is the partition function, $V$ and $N$ represent volume and the total number of particles within the volume, and the subscript means that $V$ and $N$ are kept fixed in taking derivatives.

According to the vdW theory, the entropy per particle $s$ reads as follows \cite{PhaseTrans-book,Onuki2005Dynamic}
 \begin{equation}\label{Eq-Entropy-s-perparticle}
s =  - \frac{D}{2}{k_B}\ln T + {k_B}\ln \left( {\frac{{1 - bn}}{{bn}}} \right) + {\rm{const}},
\end{equation}
where $D$ is the spatial dimension and $b$ is the covolume parameter of the vdW EOS.

In this work, we take both $k_B$ and the molecular mass at unit value, so that the $k_B$ does not appear in the
following paragraphs and the mass density $\rho$ is the same as particle number density $n$.

Combined with the generalized hydrodynamic equations Eqs. (\ref{Eq-Gns1})-(\ref{Eq-Gns3}), the relationship
between the entropy production rate and the NOEF and NOMF non-equilibrium quantities, reads as follows:
\begin{equation}\label{Eq-Entropy1}
\frac{{d{S_b}}}{{dt}} = \int {\left( {{{\pmb{\Delta }}_{3,1}^*}  \cdot \nabla  \frac{1}{T}  - \frac{1}{T} {{\pmb{\Delta }}_2^*} :\nabla {\bf{u}}} \right)d{\bf{r}}}.
\end{equation}
It can be seen that there are two source terms that directly contribute to the entropy production. The first term is NOEF (or heat flux) and the second term is NOMF (or viscous stress).

The two terms of entropy production rate are denoted by $\dot S _{NOEF}$ and $\dot S _{NOMF}$, respectively, which read
\begin{equation}\label{Eq-Entropy-rate-NOEF}
\dot S _{NOEF} = \int { {{{\pmb{\Delta }}_{3,1}^*}  \cdot \nabla  \frac{1}{T} } d{\bf{r}}},
\end{equation}
\begin{equation}\label{Eq-Entropy-rate-NOMF}
\dot S _{NOMF} = \int { { - \frac{1}{T} {{\pmb{\Delta }}_2^*} :\nabla {\bf{u}}} d{\bf{r}}}.
\end{equation}
The total entropy production rate is denoted by $\dot S_{sum}$ and it has
\begin{equation}\label{Eq-Entropy-rate-sum}
\dot S_{sum} = \dot S _{NOEF} + \dot S _{NOMF}.
\end{equation}
The contribution of the surface tension vanishes in the expression of entropy production because
the work done by the surface tension is reversible \cite{Succi2018The}.

\subsection{Numerical verification}
In order to validate the DBM for non-ideal fluids, we first assess whether the new model provides an
accurate description of the vdW EOS.

As a first test, the liquid-vapor coexistence curves at various temperatures are simulated. The computational grids are $N_x \times N_y= 200 \times 4$ with space step $\Delta x =\Delta y =0.01$, the
time-step is $\Delta t=0.0001$ and the relaxation time $\tau =0.02$. Periodic boundary conditions are adopted in both horizontal and vertical directions. The first and second order spatial derivatives are all calculated by a nine-point stencils \cite{Tiribocchi2009Hybrid,Gan2012FFT}
scheme which possesses a higher isotropy and is able to reduce spurious velocities significantly. The coefficient of surface tension is $K=5\times 10^{-5}$ and the parameters $a$ and $b$ in the EOS are chosen
as $a=\frac{9}{8}$ and $b=\frac{1}{3}$, fixing the critical point at $\rho_c = T_c=1$.

The initial conditions are
\begin{equation}\label{Eq-Initial1}
\left\{ \begin{array}{l}
{(\rho ,T,{u_x},{u_y})_L} = ({\rho _v},0.9975,0,0),\\
{(\rho ,T,{u_x},{u_y})_M} = ({\rho _l},0.9975,0,0),\\
{(\rho ,T,{u_x},{u_y})_R} = ({\rho _v},0.9975,0,0),
\end{array} \right.
\end{equation}
where the subscript ``$L$'', ``$M$'', and ``$R$'' indicate the regions $x \leq \frac{1}{4}N_x$, $\frac{1}{4}N_x< x \leq \frac{3}{4}N_x$, and $x>\frac{3}{4}N_x$, respectively, $\rho _v =0.955$ and $\rho _l =1.045$ are theoretical vapour and liquid densities, respectively, at $T=0.9975$. The temperature is dropped to $T=0.99$, until the equilibrium state is achieved.
Then the temperature drops by a small value $\Delta T=0.01$ each time the equilibrium state of the system is achieved again.

\begin{figure}[h]
\centering
  \includegraphics[height=5cm]{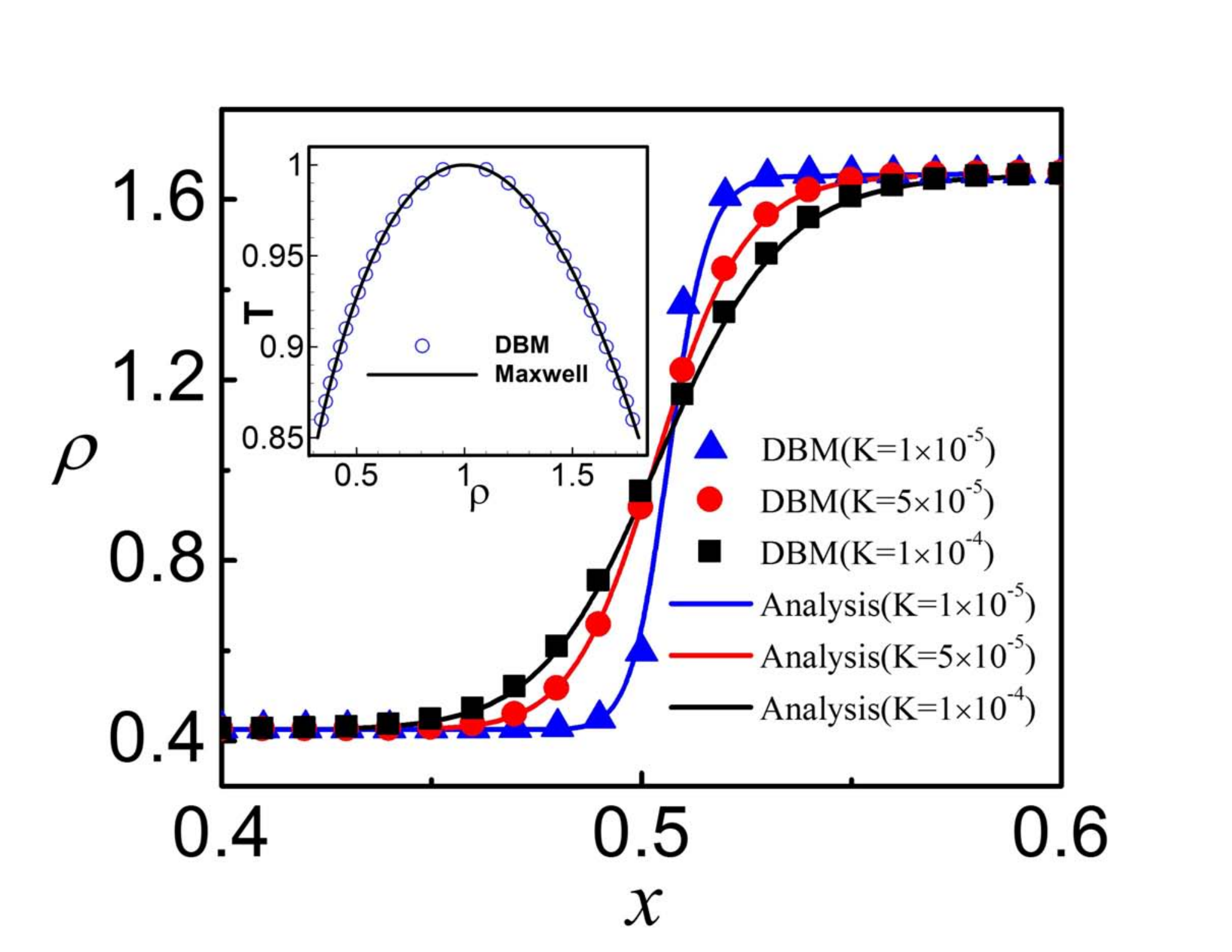}
  \caption{Equilibrium density profiles across the liquid-vapor interface for a van der Waals fluid at $T=0.90$  with three different coefficients of surface tension ($K$). The coexistence curve is also shown in the subgraph and compared with Maxwell construction. The solid lines represent analytical solutions while the symbols are results of DBM.}
  \label{fig1}
\end{figure}

Simulations go on until the temperature is reduced to $0.85$, then a series of coexistence points are obtained, as shown in the inset of Fig. \ref{fig1}. The solid line is directly calculated from van der Waals EOS using a Maxwell equal-area construction. It shows that the coexistence points simulated by the DBM  are in good agreement with the theoretical coexistence curve.
It also verifies that the DBM provides the correct vdW thermodynamics.

Figure \ref{fig1} also gives the interface density profiles at equilibrium at $T=0.90$. Three different surface tension coefficients $K$ are simulated and compared with analytical solutions \cite{Bongiorno1975Modified,Gan2012FFT}. The lines in Fig. \ref{fig1} are the analytical solutions and the symbols represent DBM simulations. It can be seen that the DBM results are in excellent agreement with the analytical solutions for the three different surface tension coefficients.

\begin{figure}[h]
\centering
  \includegraphics[height=5cm]{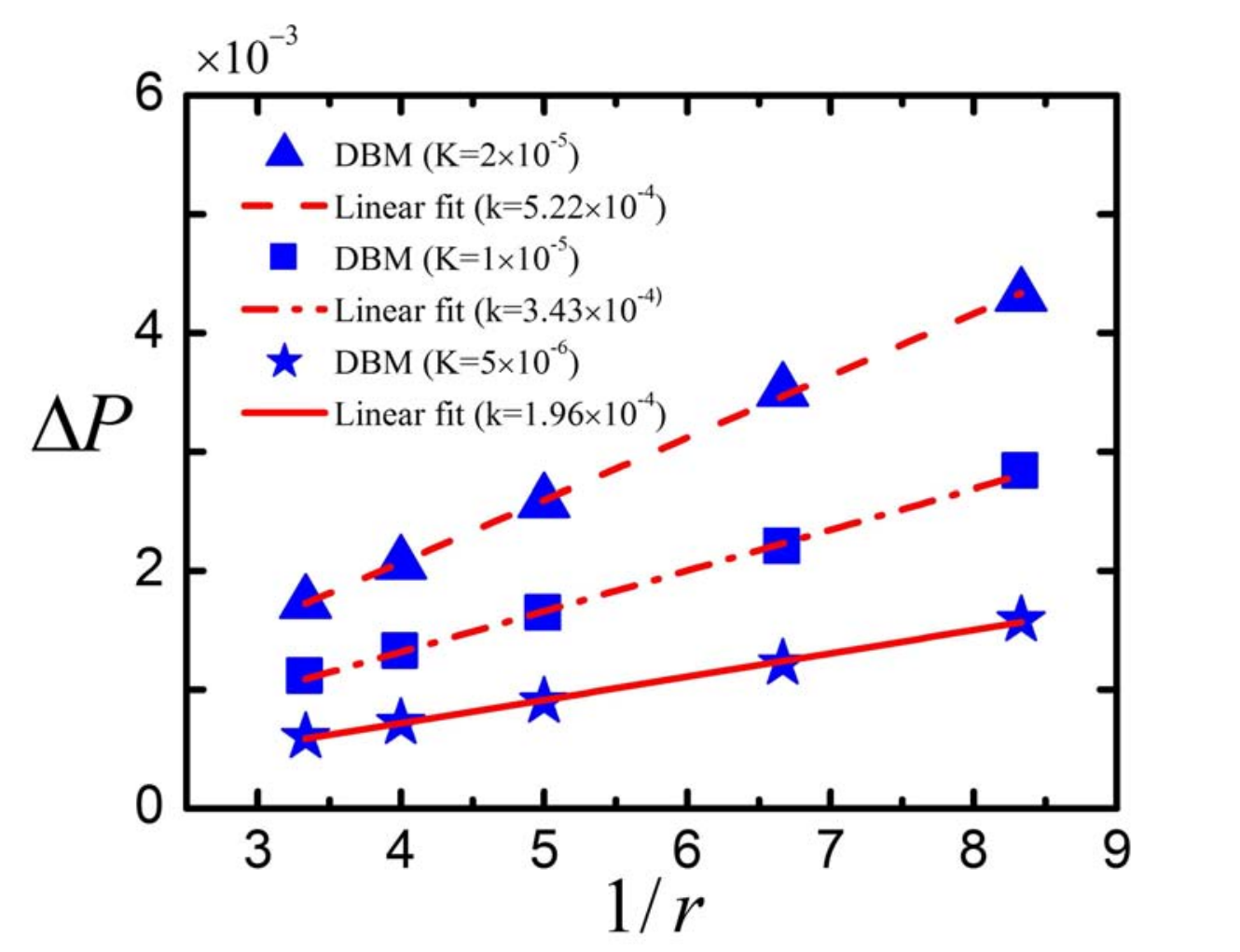}
  \caption{$\Delta P$ plotted versus $1/r$ for three different coefficients of surface tension as tests of Laplace's law. The symbols are DBM results while the lines are linear fits with slopes $k$.}
  \label{fig2}
\end{figure}

To further test our model in two-dimensions, Fig. \ref{fig2} shows the numerical validation of Laplace's law.

A circular droplet with a radius of $r$ is surrounded by its vapour phase and the initial conditions are set as follows:
\begin{equation}\label{Eq-Initial2}
\left\{ \begin{array}{l}
{(\rho ,T,{u_x},{u_y})_{{\rm{in}}}}{\kern 3pt} = (1.5865,0.92,0,0),\\
{(\rho ,T,{u_x},{u_y})_{{\rm{out}}}} = (0.4786,0.92,0,0),
\end{array} \right.
\end{equation}
where the subscript ``in'' and ``out'' indicate the regions $\sqrt{(x-L_x/2)^2+(y-L_y/2)^2} \leq r$ and $\sqrt{(x-L_x/2)^2+(y-L_y/2)^2} > r$, respectively, where $L_x$ and $L_y$ are the length and width of the computational region. The simulation region is $L_x \times L_y = 1 \times 1$, while  all the other parameters and the simulation conditions
are the same as those in Fig. \ref{fig1}.

According to Laplace law, the pressure difference $\Delta P$ between the inside and outside of the circular domain,
is proportional to the inverse radius $1/r$, when the surface tension is fixed, namely:

\begin{equation}\label{Eq-Laplace1}
\Delta P =\frac{\sigma }{r},
\end{equation}
where $\sigma$ is the surface tension.

In the simulation, three different values of $\sigma$ are used, by changing the coefficient surface tension $K$.
In Fig. \ref{fig2}, the DBM results are denoted by symbols and the lines are obtained by linear fitting. Consistently with Laplace's law, a linear relationship between the pressure difference $\Delta P$ and the
reciprocal of radius $1/r$ is obtained,

\section{Simulations and analysis}

\subsection{Isothermal and thermal phase separation}

For a single-phase fluid, such as a substance in the gas or liquid state, phase
transition and separation occur whenever the temperature suddenly drops to values consistent
with the coexistence of the two phases. Such a process is commonly called quenching and it lies at the roots of an important technology in material processing and synthesis.

Under quenching conditions, the fluid undergoes two dynamical stages: the early spinodal decomposition (SD) stage and the late stage of domain growth (DG), which correspond to the formation and the subsequent coalescence/growth
of the single-phase domains, respectively \cite{Cahn1965Phase,PhaseTrans-book,Gan2011Phase,Gan2015Discrete}.

The characteristics of the late DG stage have been extensively studied by theoretical derivations, numerical simulations, and experiments. It has been found that the characteristic domain size $R(t)$ grows in time with a power rate, $R(t)\sim t^{\alpha}$, at
 the DG stage \cite{1995OrderTheory}.  For the isothermal case, it has been found that $\alpha = 1/2$ at short times, when the growth
is mainly driven by the surface tension and diffusion and $\alpha =2/3$ for long times, when the growth is mainly driven by hydrodynamics.
DBM is at a vantage point to investigate the DG stage, since hydrodynamic models are incorporated within
the discrete Boltzmann model, in contrast to purely diffusive models \cite{Osborn1995Lattice,Gonnella1997Spinodal}.

However, the early SD stage, is comparatively less explored, possibly because
it is not clear what criteria should be adopted to draw a line between the early SD stage and the late DG stage.
Traditionally, the the critical time of the two stages is roughly marked by characteristic
domain size at the onset of the power-law regime.

A geometric criterion was provided by appealing to the so-called Minkowski function \cite{Gan2011Phase},
finding that the boundary length $L$ increases at SD stage and decreases at DG stage.
As a result, the time of maximum $L$ indicates the critical time $t_{SD}$, marking the end of the
SD stage and the beginning of the DG stage.

Subsequently, by inspecting the total TNE strength, a physical criterion was proposed to
distinguish the two stages of phase separation \cite{Gan2015Discrete}.
It was found that the total TNE strength increases at SD stage and decreases at DG stage, the maximum point corresponding
to the critical time.

In this work, we shall show that the entropy production rate $\dot S _{sum}$, can also be used
as an indicator of the transition between the SD and DG stages, the maximum of $\dot S _{sum}$ corresponding to the critical time $t_{SD}$.
\subsubsection{Simulations of two kinds of phase separation}
First, the isothermal and thermal phase separations are simulated and compared.

The initial conditions are set as follows:
\begin{equation}\label{Eq-Initial-set}
(\rho, u_x, u_y, T)=(1+\delta, 0.0, 0.0, 0.85),
\end{equation}
where $\delta$ is a random density noise with amplitude $0.01$.
The computational grids are $N_x \times N_y = 100 \times 100$, with space mesh $\Delta x = \Delta y =0.01$.
The time step is $\Delta t = 1\times 10^{-4}$ and relaxation time $\tau =0.02$.
The surface tension coefficient is $K=1 \times 10^{-5}$, the parameters in the vdW EOS
are $a=\frac{9}{8}$ and $b=\frac{1}{3}$. The spatial derivations are calculated by nine-point scheme and time is advanced by first order forward differencing.
Periodic boundary conditions are used in both directions.

For thermal case, the temperature changes freely and is solved by moment of the distribution
function at each time step. Once the distribution function
$f_{ki}$ is obtained from the evolution equation Eq. \eqref{Eq-DBM1}, the corresponding kinetic temperature is calculated through
\begin{equation}\label{Eq-SolveT}
T = \frac{1}{{2\rho }}\sum\limits_{ki} {{f_{ki}}\left( {{{\bf{v}}_{ki}} - {\bf{u}}} \right)}  \cdot \left( {{{\bf{v}}_{ki}} - {\bf{u}}} \right),
\end{equation}
where the density $\rho$ is calculated by $\rho = \sum\limits_{ki} f_{ki}$ and the macro velocity is calculated by ${\bf{u}} = \frac{1}{{\rho}}\sum\limits_{ki} {{f_{ki}} {{\bf{v}}_{ki} } } $. When simulating the isothermal case using the thermal DBM, the temperature is reset to $T=0.85$ at each time step, which
means that the temperature used to update the local equilibrium distribution function $f^{eq}_{ki}$ at each time step is fixed to be $0.85$
instead of the current value calculated from Eq. \eqref{Eq-SolveT}.

\begin{figure}[h]
\centering
  \includegraphics[height=4cm]{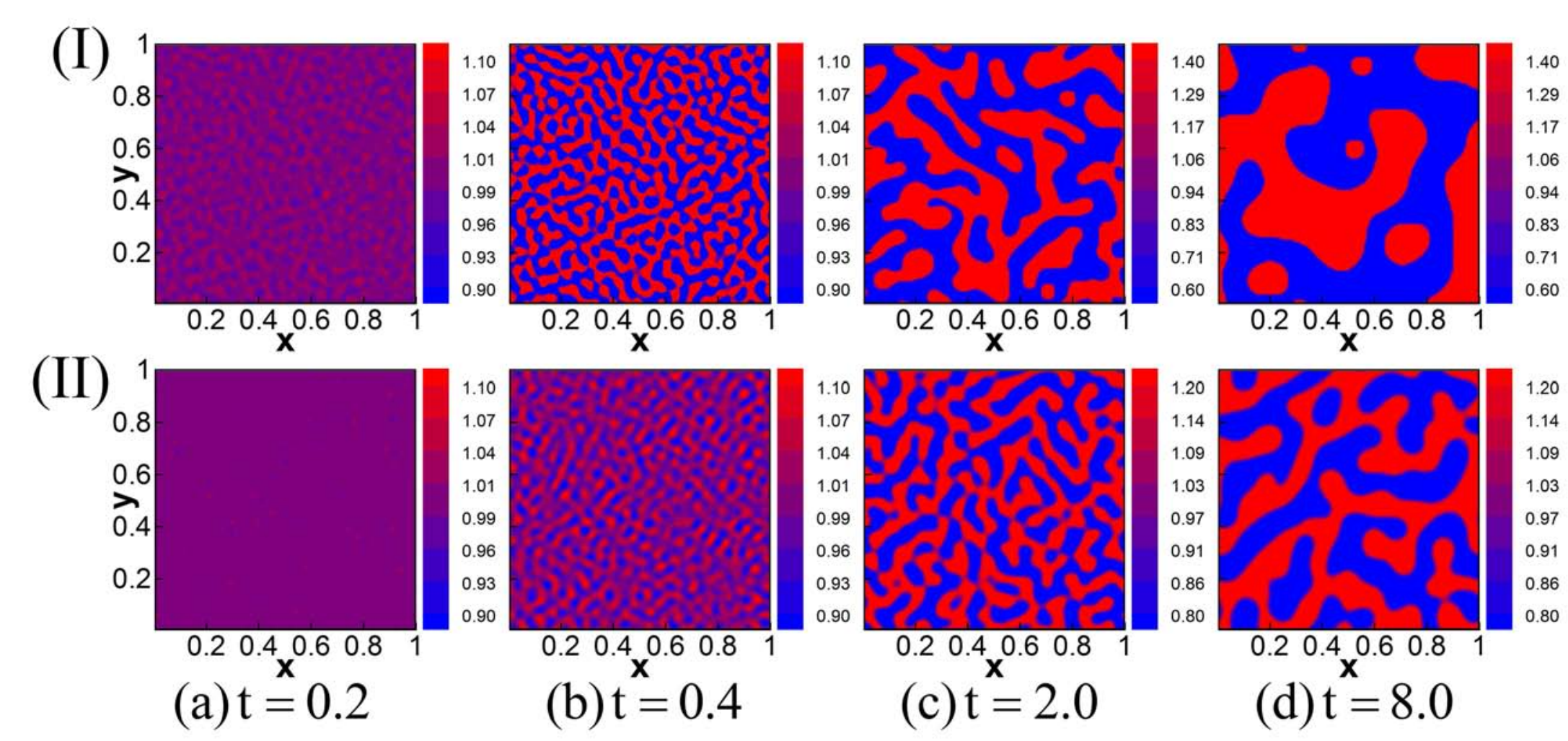}
  \caption{Density contour maps at several times for isothermal and thermal phase separation. The first line corresponds to case (I) isothermal phase separation and the second line case (II) thermal phase separation. The columns from left to right denote the density contour maps at the time (a) $t=0.2$, (b) $t=0.4$, (c) $t=2.0$, and (d) $t=8.0$, respectively.}
  \label{fig3-1}
\end{figure}

Figure \ref{fig3-1} shows the density contour maps for isothermal and thermal conditions at several typical times.
The first and second rows correspond to the isothermal case and thermal case, respectively.
The fluid separates into small regions, with higher and lower densities at $t=0.2$, then the density in the higher density region
increases and the density in the lower density region decreases.
From \ref{fig3-1} (b), one can appreciate that the liquid-vapor interfaces are clearer than those at $t=0.2$,
although there is no significant change in size of the domains of liquid or vapor.
 After the SD stage, the small domains merge and larger domains are formed under the action of surface tension.
 The characteristic domain size grows fast, as can be seen from the density contour maps
 at $t=2.0$ and $t=8.0$. The main upshot is the thermal phase separation is much slower than the isothermal one.

\subsubsection{Several criteria discriminating the two stages of phase separation}

\begin{figure}[h]
\centering
  \includegraphics[height=7cm]{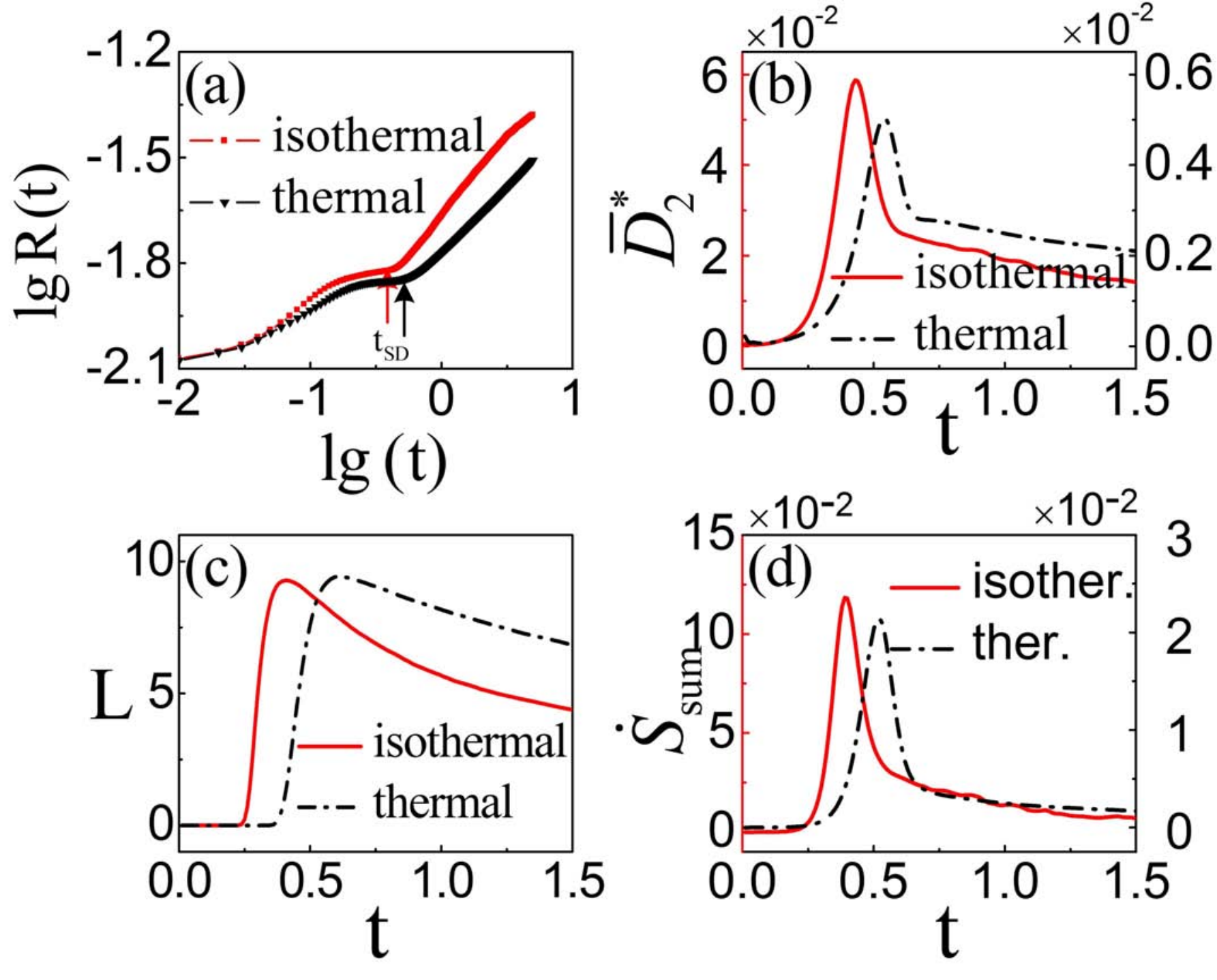}
  \caption{Four kinds of criterions to discriminate the stages of the spinodal decomposition and domain growth. (a) Characteristic domain sizes $R(t)$ in a log-log scale. (b) Non-equilibrium strengths $D^{*}_2$. (c) Boundary lengths of Morphological functional $L$. (d) The rates of total entropy production $\dot S_{sum}$.}
  \label{fig3-2}
\end{figure}

To quantitatively analyse the phase separation process, we resort to complex field analysis techniques, including statistical methods, rheological methods, and morphological methods. Figure \ref{fig3-2} gives the profiles of characteristic domain sizes $R(t)$ \cite{Gan2011Phase}, the second order thermal non-equilibrium strengths $D^*_2$ \cite{ZYD-2018arXiv}, the boundary lengths $L$ of the Minkowski functional \cite{Gan2011Phase}, and the entropy production rates $\dot S_{sum}$, as a function of time $t$.

Figure \ref{fig3-2} (a) shows the profile of $R(t)$ in a log-log scale. The characteristic domain sizes is defined as the inverse of the first moment of the structure factor $S(\mathbf{k},t)$. Generally, the structure factor is defined as the Fourier transform of the correlation function \cite{1995OrderTheory,PhaseTrans-book}. In this paper, the structure factor is calculated from the Fourier transform of order parameter as
\begin{equation}\label{eq1}
  S(k_r,t) = \frac{{\sum\limits_{\bf{k}} {\varphi ({\bf{k}},t)\varphi ( - {\bf{k}},t)} }}{{\sum\limits_{\bf{k}} 1 }},
\end{equation}
where $S(k_r,t)$ is the circularly averaged structure factor and the summation operation means averaging over a shell $k_r \leq \left| {\bf{k}} \right| < k_r + \Delta k_r$, where $\bf{k}$ represents the wave vector and $k_r$ its magnitude. In the above $\varphi ({\bf{k}},t)$ is the two-dimensional discrete Fourier transform of density fluctuation (order parameter)\cite{Corberi1998Spinodal,Corberi1999Two,Poesio2006Effects,Yokojima2002Hydrodynamic,Xu2003Phase} which reads
\begin{equation}\label{eq2}
\varphi (\mathbf{k},t) = \frac{1}{{{N_x}{N_y}{{(2\pi )}^2}}}\sum\limits_{\bf{r}} {\left[ {\rho ({\bf{r}},t) - \bar \rho (t)} \right]{e^{ i{\bf{k}} \cdot {\bf{r}}}}},
\end{equation}
where ${\bf{k}} = 2\pi \left( {\frac{m}{{{N_x}}}{{\bf{e}}_x} + \frac{n}{{{N_y}}}{{\bf{e}}_y}} \right)$ with $m=1,2,...,N_{x}$, $n=1,2,...,N_{y}$, ${\bf{e}}_x$ and ${\bf{e}}_y$ indicate two unit orthogonal vectors in the Fourier space. Then the characteristic domain size $R(t)$ is calculated by \cite{Osborn1995Lattice,Poesio2006Effects,Corberi1998Spinodal,Corberi1999Two}
\begin{equation}\label{eq3}
R(t) =2 \pi \frac{{\sum\limits_{{k_r}} {S({k_r},t)} }}{{\sum\limits_{{k_r}} {{k_r}S({k_r},t)} }}.
\end{equation}

%

The critical time $t_{SD}$ is marked by an arrow.
After the critical time, the scaling law at the DG stage builds up.
Compared with the isothermal case, the thermal phase separation shows a larger $t_{SD}$ and a smaller slope at DG stage,
which means the duration of the SD stage is longer and the growth rate of phase domain at DG stage is slower
for thermal separation than for the isothermal case. This is in accordance with the evolution of the density pattern in Fig. \ref{fig3-1}.

From previous studies, we know that the non-equilibrium strength and the boundary length of the Minkowski functional
can also be used as a criterion to discriminate the two stages of phase separation, as confirmed by inspecting
Figs. \ref{fig3-2} (b) and \ref{fig3-2} (c). The second order non-equilibrium strength $D^{*}_2$ is defined as
\begin{equation}\label{Eq-D2}
D_2^{\rm{*}} = \sqrt {{{(\Delta _{2,xx}^*)}^2} + {{(\Delta _{2,xy}^*)}^2} + {{(\Delta _{2,yy}^*)}^2}}.
\end{equation}
where the definition of non-equilibrium component $\Delta _{2,{\alpha \beta}}^*$ is given in Eq. \eqref{Eq-Deltaxing-m}. The Minkowski functionals are a set of statistical indicators first proposed by Minkowski \cite{Minkowski1903}.
According to integral geometrical criteria, all properties of a $n$-dimensional convex set which satisfy
translational invariance and additivity, can be fully described by $n+1$ functionals, known as Minkowski functionals \cite{Xu2009Morphological,Gan2011Phase}.
For the  two-dimensional case, the three functionals are the fractional area, the boundary length, and the Euler characteristic \cite{Gan2011Phase}.

To obtain the Minkowski functionals of the density map, we first choose a density threshold $\rho_{th}$. The computational domain is then divided into high-density regions, where $\rho \geq \rho_{th}$, and low-density regions
where $\rho < \rho_{th}$.
The boundary length $L$ is defined as the sum of the dividing lines between high-density and the low-density regions, i.e.
the length of the interface.
Both non-equilibrium strength and boundary length increase with time at the SD stage (roughening)
and decrease in the DG stage (coarsening), the maximum points corresponding to the critical times $t_{SD}$.

The non-equilibrium strength provides a physical criterion, while the boundary length of Minkowski functional $L$
provides a geometrical one.
The combined resort to these two criteria facilitates the identification of the critical time of the SD stage and the DG stage.

The profiles of total entropy production rate $\dot S _{sum}$ are plotted in Fig. \ref{fig3-2} (d).
We can see that the entropy production rate increases with time in the SD stage and decreases with time in the DG stage. It attains a maximum value when the SD stage ends and the DG stage begins, hence providing a further
physical criterion for the two stages of phase separation.

It should be noted that the values of $t_{SD}$, given by the different criteria, may not be exactly the same.
This is physically reasonable, as it reflects the different characteristic of the complex flows as viewn from different perspectives.
Nevertheless, they all lead to the  same conclusion: isothermal phase separation is faster than the thermal one. In addition, we see that the non-equilibrium strength of thermal phase separation is weaker
and the entropy production rate of thermal phase separation is lower than the isothermal case, see Figs. \ref{fig3-2} (b) and \ref{fig3-2} (d), respectively.

\subsection{Effects of heat conduction}
\subsubsection{Thermal phase separation with different heat conductivities}
To examine the effects of heat conduction, we simulate the thermal phase separation with several different heat conductivities.
In the usual DBM based on a single-relaxation time model, the coefficients of viscosity $\mu$ and heat conductivity
$\kappa$ are bound together, namely $\mu =\tau \rho T$ and $\kappa = c_p \tau \rho T$. The Prandtl number, $Pr=c_p \mu/ \kappa$, is fixed and consequently
the effects of viscosity and heat conduction are usually considered together.

However, in the multiphase flow DBM, by introducing the $C_q$ in the extra term \cite{Gonnella2007Lattice,Gan2015Discrete}, the $Pr$ is adjustable.
Under the fixed viscosity coefficient, different heat conductivities can be obtained, by changing the value of Prandtl
under the fixed relaxation time $\tau$. Thus, the heat conductivity can be represented by $1/Pr$.

All the simulation conditions here are the same with those of thermal phase separation in Fig. \ref{fig3-1} except for the value of heat conductivity (or Pr). Different heat conductivities are obtained by changing the value of $C_q$ in Eq. \eqref{Eq-Iki}. The relationship between the $C_q$ and the $Pr$ refers to previous literatures \cite{Gonnella2007Lattice,Gan2015Discrete}.

\begin{figure}[h]
\centering
  \includegraphics[height=6cm]{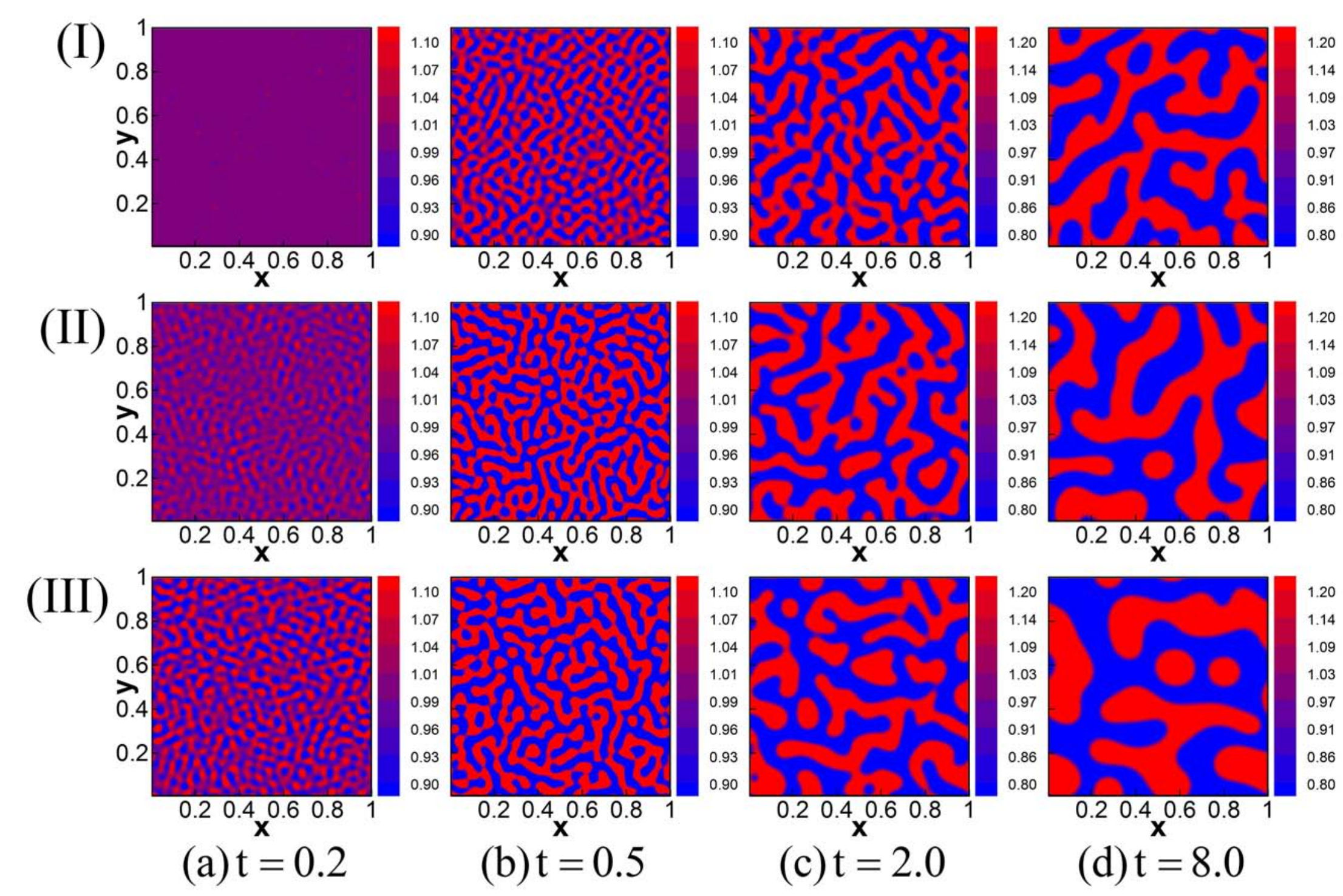}
  \caption{Density contour maps at several times for three different heat conductivities (represented by $1/Pr$) with the same viscosity coefficient. The first line corresponds to case (I) $Pr$=1.0, the second line case (II) $Pr=0.5$, and the third line case (III) $Pr=0.2$. The columns from left to right denote the density contour maps at the time (a) $t=0.2$, (b) $t=0.5$, (c) $t=2.0$, and (d) $t=8.0$, respectively.}
  \label{fig4-1}
\end{figure}

Figure \ref{fig4-1} shows a time sequence of density contour maps for three different values of $Pr$.
The three lines from top to bottom correspond to $Pr=1.0$, $Pr=0.5$, and $Pr=0.2$, respectively.
The four columns from left to right correspond to the snapshots at time $t=0.2$, $t=0.5$, $t=2.0$, and $t=8.0$, respectively.
From the contour maps of density, we see that the smallest $Pr$, the fastest the phase-separation.
For a fixed viscosity coefficient, a smaller $Pr$ corresponds to a larger heat conductivity, hence
we conclude that heat conductivity accelerates the evolution of thermal phase-separation, by
promoting the formation of the liquid-vapor phase interfaces in the SD stage and facilitating the merge of small domains in the DG stage.

\begin{figure}[h]
\centering
  \includegraphics[height=6cm]{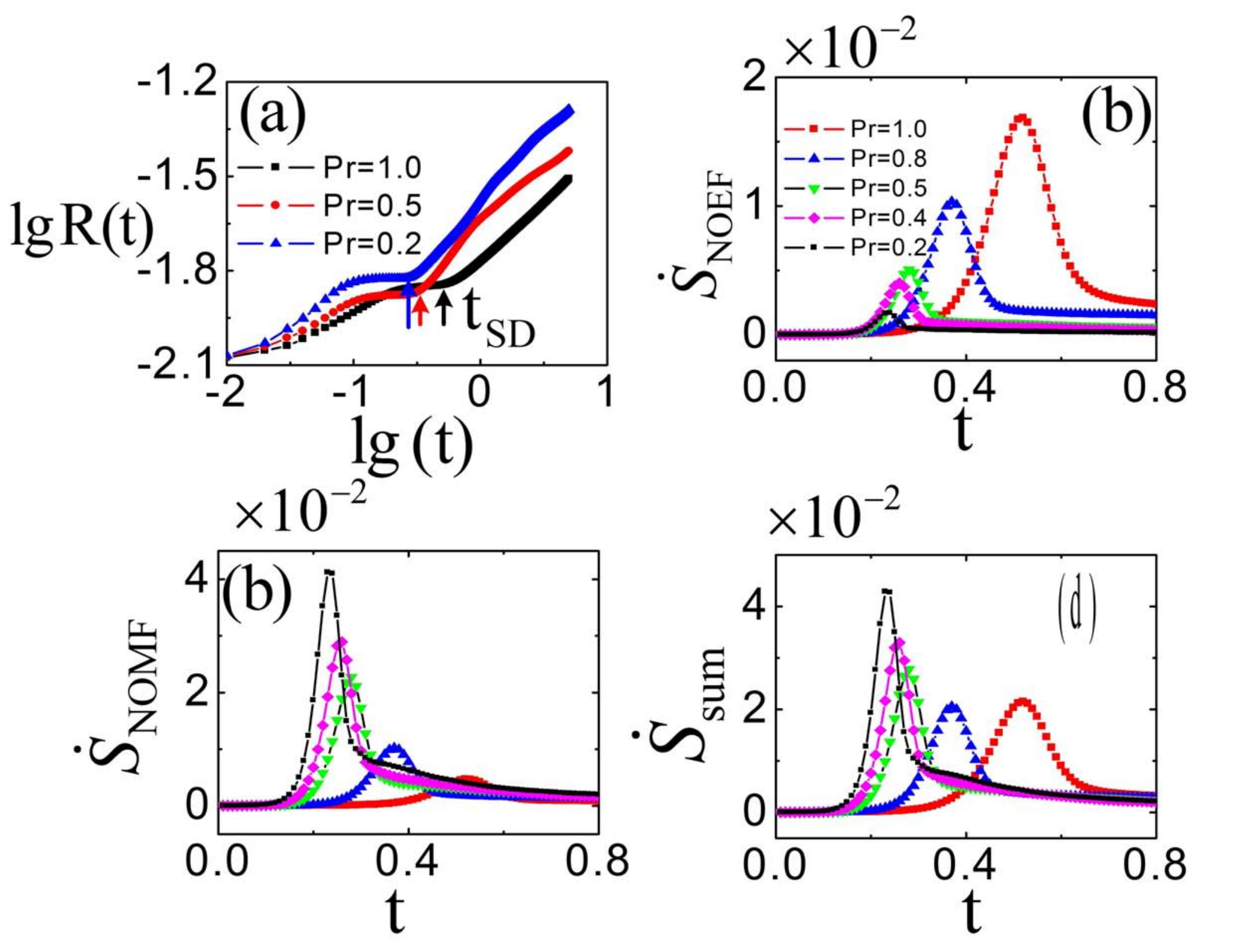}
  \caption{The profiles of characteristic domain sizes and the rates of entropy production for different Prandtl numbers. (a) The characteristic domain sizes $R(t)$ in a log-log scale. (b) The entropy production rate of NOEF $\dot S_{NOEF}$. (c) The entropy production rate of NOMF $\dot S_{NOMF}$. (d) The total entropy production rate $\dot S_{sum}$. The last three subgraphs share the same legend shown in the subgraph (b).}
  \label{fig4-2}
\end{figure}

To quantify the characteristics of the evolution of thermal phase separation with different heat conductivities, the profiles of the characteristic domain size $R(t)$ and the entropy production rates, $\dot S_{NOEF}$, $\dot S_{NOMF}$, and $\dot S_{sum}$, as a function of time for several different Prandtl numbers, are plotted in Fig. \ref{fig4-2}.

The observable $R(t)$ provides an approximate criterion to distinguish the two stages of thermal phase separation.
First, $R(t)$ increases and reaches a plateau, which remains until the end of the SD stage.
Then, phase-separation enters the DG stage and $R(t)$ grows in the form of a power law.
The end of the plateau corresponds to the critical time $t_{SD}$, which is marked by arrow in Fig. \ref{fig4-2} (a).
The larger $Pr$ (or the smaller the heat conductivity), the longer $t_{SD}$ is.
This means that heat conduction accelerates the spinodal decomposition process, thus shortening
the duration of the SD stage and speed up thermal phase separation.
The same conclusions can be drawn by inspecting the maxima of entropy production rates in Figs. \ref{fig4-2} (b) - \ref{fig4-2} (d).

From Fig. \ref{fig4-2} (b), we see that the profiles of $\dot S _{NOEF}$ shift leftwards and the amplitudes decrease
with increasing heat conductivity.
This indicates that the effect of heat conduction is to reduce the entropy production rate of NOEF.
The profiles of $\dot S _{NOMF}$ in Fig. \ref{fig4-2} (c), also shift leftwards, while the amplitude increases with the heat conductivity.
In addition, it is observed that the NOMF entropy production occurs mainly in a very short period of time, during the early stage of phase-separation, and
it almost stops at later times. The larger the heat conductivity, the more concentrated the NOMF entropy production.
The NOEF entropy production shares similar characteristic, though it is not significant, due to its lower amplitude
when the heat conductivity is larger.

The total entropy production rate $\dot S_{sum}$, is the sum of the two kinds of entropy production
rates, $\dot S_{NOEF}$ and $\dot S_{NOMF}$. Since $\dot S_{NOEF}$ decreases, while $\dot S_{NOMF}$ increases with the increase
of heat conductivity, the effects of the heat conductivity cancel out on the behaviour of total entropy production rate.
From Fig. \ref{fig4-2} (d), we see that the change of the amplitude of $\dot S_{sum}$ is marginal at $Pr \sim 1$,
but the increase of amplitude is significant with the increase of heat conductivity, for $Pr<0.8$.

\subsubsection{Effects of heat conduction on the SD stage and the entropy production rates}
Now, we run additional simulations with different heat conductivities and further investigate the characteristic
duration time of the SD stage, $t_{SD}$, and the entropy production rates, $\dot S_{NOMF}$ and $\dot S_{NOEF}$, under the effect of heat conductivity.

\begin{figure}[h]
\centering
  \includegraphics[height=2.5cm]{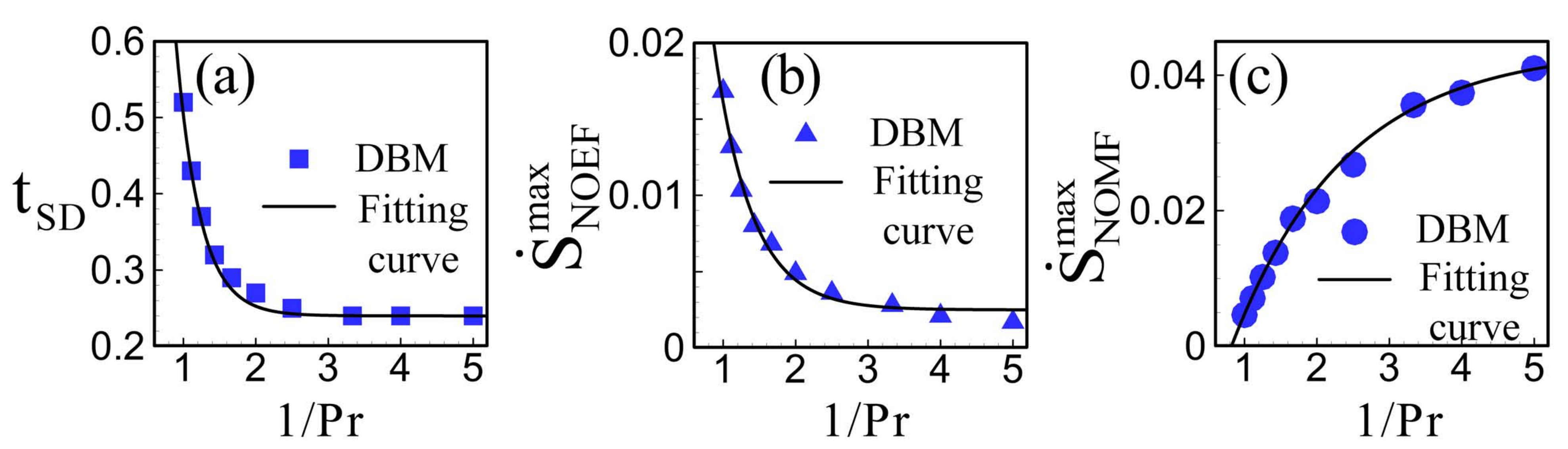}
  \caption{The profiles of the critical time $t_{SD}$ and the maximum entropy production rates as functions of heat conductivity, (a) the duration time of the SD stage $t_{SD}$, (b) the maximum entropy production rate of NOEF $\dot S^{max} _{NOEF}$, and (c) the maximum entropy production rate of NOMF $\dot S^{max} _{NOMF}$. The symbols are results of DBM and the solid lines are fitting curves.}
  \label{fig4-3}
\end{figure}

Figure \ref{fig4-3} (a) gives the results of $t_{SD}$, simulated by DBM with different values of heat conductivity (represented by $1/Pr$). When the heat conductivity changes in the range of $1 \leq 1/Pr \leq 5$, the duration time of SD stage $t_{SD}$ decrease with the increase of the heat conductivity. The relationship between $t_{SD}$ and $1/Pr$ can be fitted by a decaying exponential function
 \begin{equation}\label{Eq-tSD-fitting1}
t_{SD}=\mathrm{C_1} \exp (-\mathrm{C_2}/ Pr)+\mathrm{C_0},
\end{equation}
where $\mathrm{C_1} = 5.46$, $\mathrm{C_2}=3.03$, and $\mathrm{C_0}=0.25$. $\mathrm{C_0}$ is the value of $t_{SD}$ in the limit of zero Prandtl number, i.e. infinite thermal conductivity. Note that $\mathrm{C_0}$ is non-zero because the SD stage would not disappear even in the limit of infinite thermal conductivity. From Fig. \ref{fig4-3} (a) we can see the duration time of SD stage barely changes when $1/Pr > 3$, the minimum value of $t_{SD}$ is about $0.25$. Oppositely, as the heat conductivity approaches zero the duration time of SD stage $t_{SD}$ approaches $\mathrm{C_0}+\mathrm{C_1}$. Under the effect of heat conductivity, the difference between the maximum and minimum value of $t_{SD}$ is about $20$ times. Finally, $\mathrm{C_2} \sim 3$, is basically the Prandtl number below which exponential decay becomes manifest. This expression confirms the previous conclusion that heat conduction helps accelerating the SD stage of thermal phase separation and duration of SD stage, $t_{SD}$, decreases exponentially with heat conductivity.

Next, we examine the effects of heat conduction on the entropy production rate $\dot S _{NOEF}$ and $\dot S _{NOMF}$.
The  peak values of $\dot S _{NOEF}$ and $\dot S _{NOMF}$ are taken to represent the corresponding amplitudes of
the entropy production rates. They are indicated by $\dot S^{max}_{NOEF}$ and $\dot S^{max}_{NOMF}$, respectively.

\begin{figure}[h]
\centering
  \includegraphics[height=4cm]{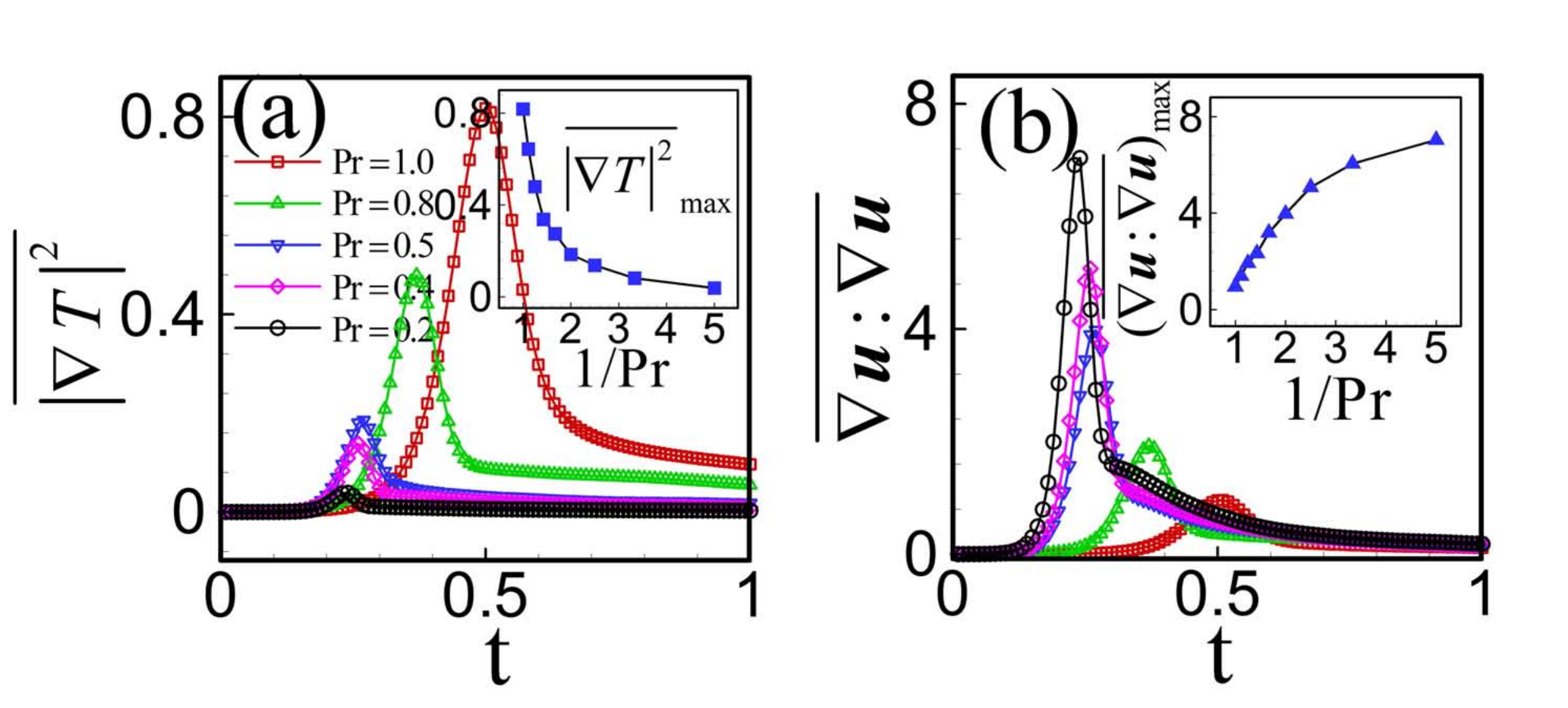}
  \caption{The profiles of the average values of temperature gradient and velocity gradient for different Prandtl numbers, (a) the average value of temperature gradient $\overline {{{\left| {\nabla T} \right|}^2}} $ and (b) the average value of velocity gradient $\overline {\nabla {\bf{u}}:\nabla {\bf{u}}} $.}
  \label{fig4-5}
\end{figure}

 Figures \ref{fig4-3} (b) shows that $\dot S^{max}_{NOEF}$ decrease with increasing heat conductivity, while Fig. \ref{fig4-3} (c) shows that $\dot S^{max}_{NOMF}$ increase with the increase of heat conductivity. These can be explained from the temperature gradient and the velocity gradient in the flow field, respectively. With the increase of the heat conductivity, the heat conduction effect is strengthened, resulting in a more uniform temperature field. From the expression of the $\dot S_{NOEF}$ in Eq. \eqref{Eq-Entropy-rate-NOEF}, we can learn that the entropy production of NOEF relies on two aspects, the NOEF $\pmb{\Delta}^*_{3,1}$ and the temperature gradient $\nabla T$. In fact, in the hydrodynamic limit, $\pmb{\Delta}^*_{3,1} \approx -\kappa \nabla T$ where $\kappa$ indicates the heat conductivity. This means that the temperature gradient is the dominant factor affecting the entropy production rate $\dot S_{NOEF}$, and the integral part in $\dot S_{NOEF}$ can be approximated by $\kappa \left| {\nabla T} \right|^2 /T^2$. The profile of the average value of $\left| {\nabla T} \right|^2 $, denoted by $\overline {{{\left| {\nabla T} \right|}^2}} $, is plotted in Fig. \ref{fig4-5} (a), from which we can see that the characteristics of $\overline {{{\left| {\nabla T} \right|}^2}} $ under various Prandtl numbers are very similar to those of $\dot S_{NOEF}$ in Fig. \ref{fig4-2} (b). Besides, the peak values of $\overline {{{\left| {\nabla T} \right|}^2}} $ as a function of $1/Pr$ are plotted in the inset of Fig. \ref{fig4-5} (a), and it is very similar with the profile of $\dot S^{max}_{NOEF}$ in Fig. \ref{fig4-3} (b). This confirms the previous explanation. Increasing heat conductivity contributes to the decrease of $\dot S_{NOEF}$ by smoothing the temperature distribution in the flow field.

 Besides, from Fig. \ref{fig4-3} (b) we can see that the relationship between the $\dot S^{max}_{NOEF}$ and $1/Pr$ can be fitted by an exponential function, like the one in Eq. \eqref{Eq-tSD-fitting1}. The fitting coefficients are $\mathrm{C_1} = 0.094$, $\mathrm{C_2}=1.93$, and $\mathrm{C_0}=0.0026$. The amplitude of the entropy production rate of NOEF decreases exponentially with heat conductivity. The minimum value of $\dot S^{max}_{NOEF}$ is close to zero when the heat conductivity tends to infinity. The reason is that the temperature gradient disappears when the heat conductivity tends to infinity. The maximum value of $\dot S^{max}_{NOEF}$ is about $0.094$ when the heat conductivity approaches zero. The quantity $1/\mathrm{C_2}=0.52$ is the typical scale of heat conductivity above which the exponential decay of $\dot S^{max}_{NOEF}$ is manifest.

 On the other hand, since the heat conduction speeds up the process of the thermal phase separation, it promotes the mutual motion between different phase domains and therefore increases the velocity gradient in the flow field. From the expression of the $\dot S_{NOMF}$ in Eq. \eqref{Eq-Entropy-rate-NOMF}, we can see that the entropy production of NOMF depends on two quantities, the NOMF $\pmb{\Delta}^*_{2}$ and the velocity gradient $ \nabla \mathbf{u}$. In the hydrodynamic limit, one has $\pmb{\Delta}^*_{2} \approx -\mu \left[ {\nabla {\bf{u}} + {{(\nabla {\bf{u}})}^T} - (\nabla  \cdot {\bf{u}}){\bf{I}}} \right]$ where $\mu$ is coefficient of viscosity. It can be seen that the velocity gradient is the dominant factor affecting the entropy production rate $\dot S_{NOMF}$, and the integral part in $\dot S_{NOMF}$ can be approximated by $\mu \left[ {\nabla {\bf{u}}:\nabla {\bf{u}} + {{(\nabla {\bf{u}})}^T}:\nabla {\bf{u}} - {{\left| {\nabla  \cdot {\bf{u}}} \right|}^2}} \right]/T$. In addition, we find that the characteristics of $ {\nabla {\bf{u}}:\nabla {\bf{u}} + {{(\nabla {\bf{u}})}^T}:\nabla {\bf{u}} - {{\left| {\nabla  \cdot {\bf{u}}} \right|}^2}} $ and $\nabla {\bf{u}}:\nabla {\bf{u}}$ are very similar to each other.
For simplicity, we only consider the characteristics of the latter.  Fig. \ref{fig4-5} (b) reports the profiles of $\overline {\nabla {\bf{u}}:\nabla {\bf{u}}} $, which is defined as the average value of $\nabla {\bf{u}}:\nabla {\bf{u}}$. We can see the profiles under different Prandtl numbers are very similar as those of $\dot S_{NOMF}$ in Fig. \ref{fig4-2} (c). The peak values of $\overline {\nabla {\bf{u}}:\nabla {\bf{u}}} $ as a function of $1/Pr$ are also plotted in the inset of Fig. \ref{fig4-5} (b), which shows an increase of the peak values at increasing heat conductivity, similar to the increase of $\dot S^{max}_{NOMF}$ in Fig. \ref{fig4-3} (c). This shows that increasing heat conductivity contributes to the increase of $\dot S_{NOMF}$ by promoting the relative motion between different phase domains and increasing the velocity gradient in the flow field.

In addition, as shown in Fig. \ref{fig4-3} (c), the relationship between the $\dot S^{max}_{NOMF}$ and $1/Pr$ is well fitted by an increasing exponential:
  \begin{equation}\label{Eq-SNOMF-fitting1}
\dot S_{NOMF}=-\mathrm{C_1} \exp (-\mathrm{C_2}/ Pr)+\mathrm{C_0},
\end{equation}
  where $\mathrm{C_1} = 0.074$, $\mathrm{C_2}=0.634$, and $\mathrm{C_0}=0.044$. This shows that the amplitude of the entropy production rate of NOMF increases exponentially with heat conductivity. The maximum value of $\dot S^{max}_{NOMF}$ is about 0.044, corresponding to the limit of infinite thermal conductivity. However, the value of $\dot S^{max}_{NOMF}$ may be negative when $1/Pr <1$ which is unreasonable since the entropy production can not be negative. So this fitting result is only applicable when $1/Pr >1$. The typical scale of heat conductivity for $\dot S^{max}_{NOMF}$ is $1/\mathrm{C_2}=1.58$ which is significantly larger than that for $\dot S^{max}_{NOEF}$.

\subsection{Effects of viscosity}
\subsubsection{Thermal phase separation with different viscosity coefficients}
 In this section, we study the effects of viscosity on the thermal phase separation. The simulation conditions are the same with those in Fig. \ref{fig3-1} (II). Various viscosity coefficients are obtained by changing the relaxation time $\tau$ since $\mu =\tau \rho T$.

\begin{figure}[h]
\centering
  \includegraphics[height=6cm]{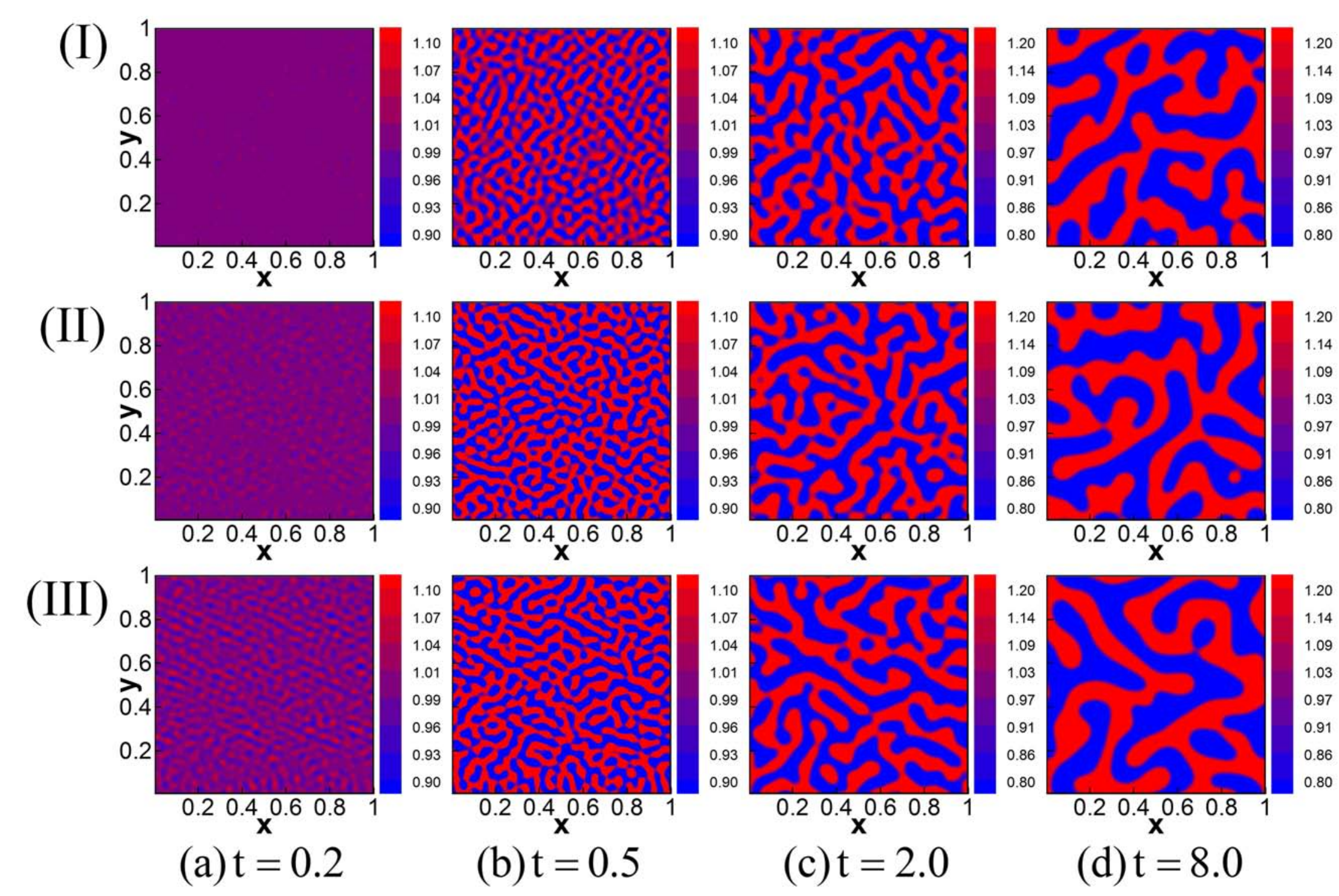}
  \caption{Density contour maps at several times for different coefficients of viscosity (represented by $\tau$) with the same heat conductivity. The first line corresponds to case (I) $\tau=0.02$ and Pr=1.0, the second line case (II) $\tau=0.016$ and Pr=0.8 and the third line case (III) $\tau=0.01$ and Pr=0.5. The columns from left to right denote the density contour maps at the time (a) $t=0.2$, (b) $t=0.5$, (c) $t=2.0$, and (d) $t=8.0$, respectively.}
  \label{fig5-1}
\end{figure}

 To keep the heat conductivity
 $\kappa = \mu c_p /Pr$ fixed, the Prandtl number also needs to change with the relaxation time.
 Figure \ref{fig5-1} shows the density contour maps for three different values of $\tau$ at several typical time instants.
 The three lines from top to bottom correspond to the cases with (I) $\tau=0.02$ and Pr$=1.0$, (II) $\tau=0.016$ and Pr$=0.8$ and (III) $\tau=0.01$ and Pr$=0.5$, respectively. The four columns from left to right correspond to the snapshots at time (a) $t=0.2$, (b) $t=0.5$, (c) $=2.0$, and (d) $t=8.0$, respectively.

 From the density contour maps at $t=0.2$, we see that a smaller viscosity coefficient (represented by $\tau$) corresponds to a sharper interface, which indicates that the weaker the viscosity, the faster the thermal phase separation process in the SD stage.
 Besides, in the DG stage, a smaller viscosity coefficient corresponds to a larger area of phase domain, which can
 be seen from the contour maps at $t=2.0$ and $t=8.0$, though this difference is not very significant.

\begin{figure}[h]
\centering
  \includegraphics[height=6 cm]{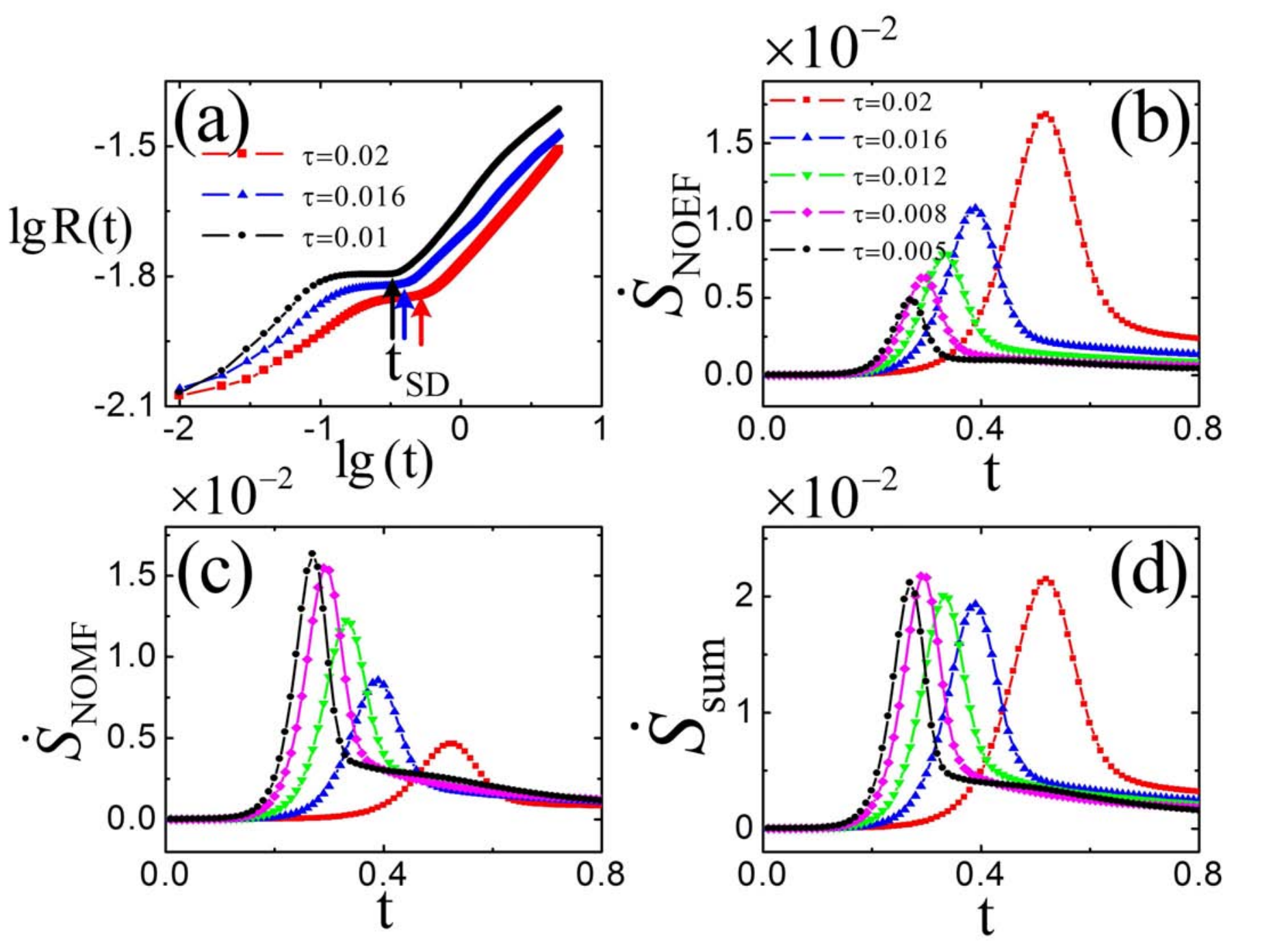}
  \caption{The profiles of characteristic domain sizes and the rates of entropy production for different viscosity coefficients. (a) The characteristic domain sizes $R(t)$ in a log-log scale. (b) The entropy production rate of NOEF $\dot S_{NOEF}$. (c) The entropy production rate of NOMF $\dot S_{NOMF}$. (d) The total entropy production rate $\dot S_{sum}$. The last three subgraphs share the same legend shown in the subgraph (b). }
  \label{fig5-2}
\end{figure}

Figure \ref{fig5-2} (a) shows the profiles of characteristic domain size $R(t)$ in a log-log scale for three different viscosity coefficients. The duration times of the SD stage, $t_{SD}$, are marked by arrows from which we know the roles of viscosity are to inhibit the formation of the phase domains and to prolong the SD stage. The rate of the phase domain growth can be represented by the slope of the $\log R(t)$ from which we can see the viscosity has little effect on the rate of separation at the DG stage. The reason is that viscosity coefficients are all much higher in those simulation conditions, the effect of viscosity changing on the domain growth is not significant. According to the previous study we have learned that, at DG stage, the growth rate reads $R(t) \sim t ^{1/2}$ for higher viscosity and $R(t) \sim t ^{2/3}$ for lower viscosity \cite{1995OrderTheory,Gan2011Phase,Osborn1995Lattice}. The power exponent is not sensitive to the viscosity coefficient within a certain range.  Although this conclusion is for isothermal conditions, we find that similar features exist in thermal phase separation.

The profiles of entropy production rates, $\dot S_{NOEF}$, $\dot S_{NOMF}$, and $\dot S_{sum}$, for different viscosity coefficients are shown in Figs. \ref{fig5-2} (b), \ref{fig5-2} (c), and \ref{fig5-2} (d), respectively. The maximum points of entropy production rates indicate the critical times $t_{SD}$ which are consistent with those in Fig. \ref{fig5-2} (a). The larger the viscosity coefficient is, the longer the spinodal decomposition stage lasts.

 From Fig. \ref{fig5-2} (b) we can see that the profile of $\dot S _{NOEF}$ shifts to the right and the amplitude increases with the increase of viscosity coefficient, from which we know that the role of viscosity is to increase the entropy production rate of NOEF. The profiles of $\dot S_{NOMF}$ for several different kinds of viscosity coefficients are shown in Fig. \ref{fig5-2} (c). With the increase of viscosity coefficient, the amplitude of the entropy production rate decrease. This indicates that the role of viscosity is to decrease the entropy production rate of NOMF. Besides, it can be found that the entropy production occurs mainly in the early stage of the thermal phase separation. This can be explained as follow, at the late stage, the surface tension plays a leading role which has no contribution to the entropy production. The smaller the viscosity coefficient, the more concentrated the entropy production because the effect of viscosity is nearly negligible compared to the surface tension at the late stage. The total entropy production rate $\dot S_{sum}$ is the sum of the two parts, $\dot S_{NOEF}$ and $\dot S_{NOMF}$.

 From Fig. \ref{fig5-2} (d) we can see that there is little difference in the amplitudes of total entropy production rate $\dot S_{sum}$ between the cases with different viscosity coefficients. Since the amplitude of $\dot S_{NOEF}$ increase while the amplitude of $\dot S_{NOMF}$ decrease with the increase of viscosity coefficient, they cancel each other out. So the effect of viscosity on the amplitude of total entropy production rate is not significant.
\subsubsection{Effects of viscosity on the SD stage and the entropy production rates}

To further investigate the effect of viscosity on thermal phase separation, we run more simulation cases with viscosity coefficient ($\tau$) varying within the range $[0.005,0.02]$.

\begin{figure}[h]
\centering
  \includegraphics[height=3cm]{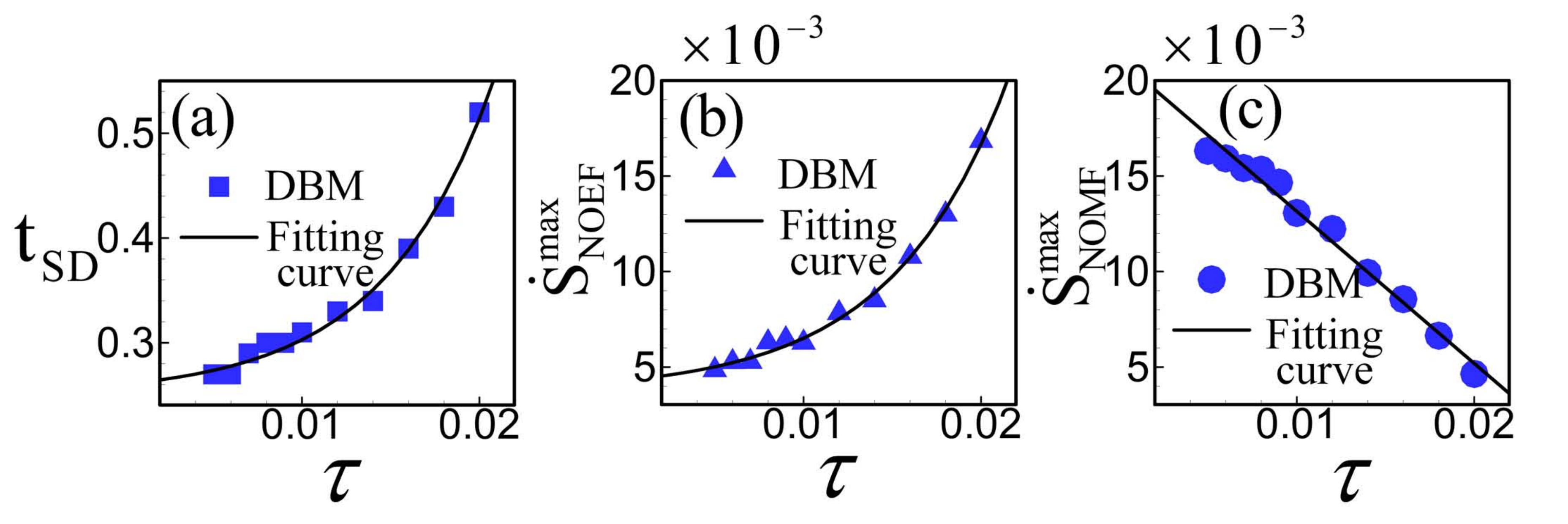}
  \caption{The profiles of critical time $t_{SD}$ and the amplitudes of entropy production rates as functions of viscosity coefficient. (a) the duration time of the SD stage $t_{SD}$, (b) the maximum entropy production rate of NOEF $\dot S^{max} _{NOEF}$, (c) the maximum entropy production rate of NOMF $\dot S^{max} _{NOMF}$. The symbols are results of DBM and the solid lines are fitting curves.}
  \label{fig5-3}
\end{figure}

The relationship between the duration time of the SD stage, $t_{SD}$, and viscosity coefficient is shown in Fig. \ref{fig5-3} (a). The symbols are DBM results, from which we can see that the duration time of the SD stage increases with the increase of viscosity coefficient. The relationship between $t_{SD}$ and viscosity coefficient can be fitted by the following exponential function:
\begin{equation}\label{Eq-tSD-fitting2}
t_{SD}= \mathrm{C_1} \exp (\mathrm{C_2} \tau)+ \mathrm{C_0} ,
\end{equation}
where $\mathrm{C_1} = 0.0106$, $\mathrm{C_2}=160.77$, and $\mathrm{C_0}=0.25$. The solid line in Fig. \ref{fig5-3} (a) is the fitting result, from which we conclude that the duration time of SD stage $t_{SD}$ increases exponentially with viscosity coefficient within the range of simulation parameters. $\mathrm{C_0}+\mathrm{C_1}$ is the value of $t_{SD}$ when the coefficient of viscosity tends to zero, which is almost equal to the value of $t_{SD}$ when the heat conductivity tends to infinity. As the viscosity coefficient decreases, the rate of decay of $t_{SD}$ decreases gradually and finally $t_{SD}$ approaches $0.25$. Oppositely, as viscosity coefficient increases, the duration time of SD stage increases significantly. $1/\mathrm{C_2}=0.0062$ is the typical scale of viscosity coefficient above which the exponential decay of $t_{SD}$ becomes manifest and $\mathrm{C_1}=0.0106$ is the amplitude of the exponential growth.

Next, we examine the effects of viscosity on entropy production rates, $\dot S _{NOEF}$ and $\dot S _{NOMF}$. The peak values $\dot S^{max}_{NOEF}$ and $\dot S^{max}_{NOMF}$ are taken as representative of the amplitudes of the entropy production rate and are plotted in Figs. \ref{fig5-3} (b) and \ref{fig5-3} (c), respectively, as functions of $\tau$. The symbols are simulation results where both $\dot S^{max}_{NOEF}$ and $\dot S^{max}_{NOMF}$ are given in units of $10^{-2}$. It is apparent that $\dot S^{max}_{NOEF}$ increases while $\dot S^{max}_{NOMF}$ decreases with the increase of viscosity coefficient.

\begin{figure}[h]
\centering
  \includegraphics[height=4cm]{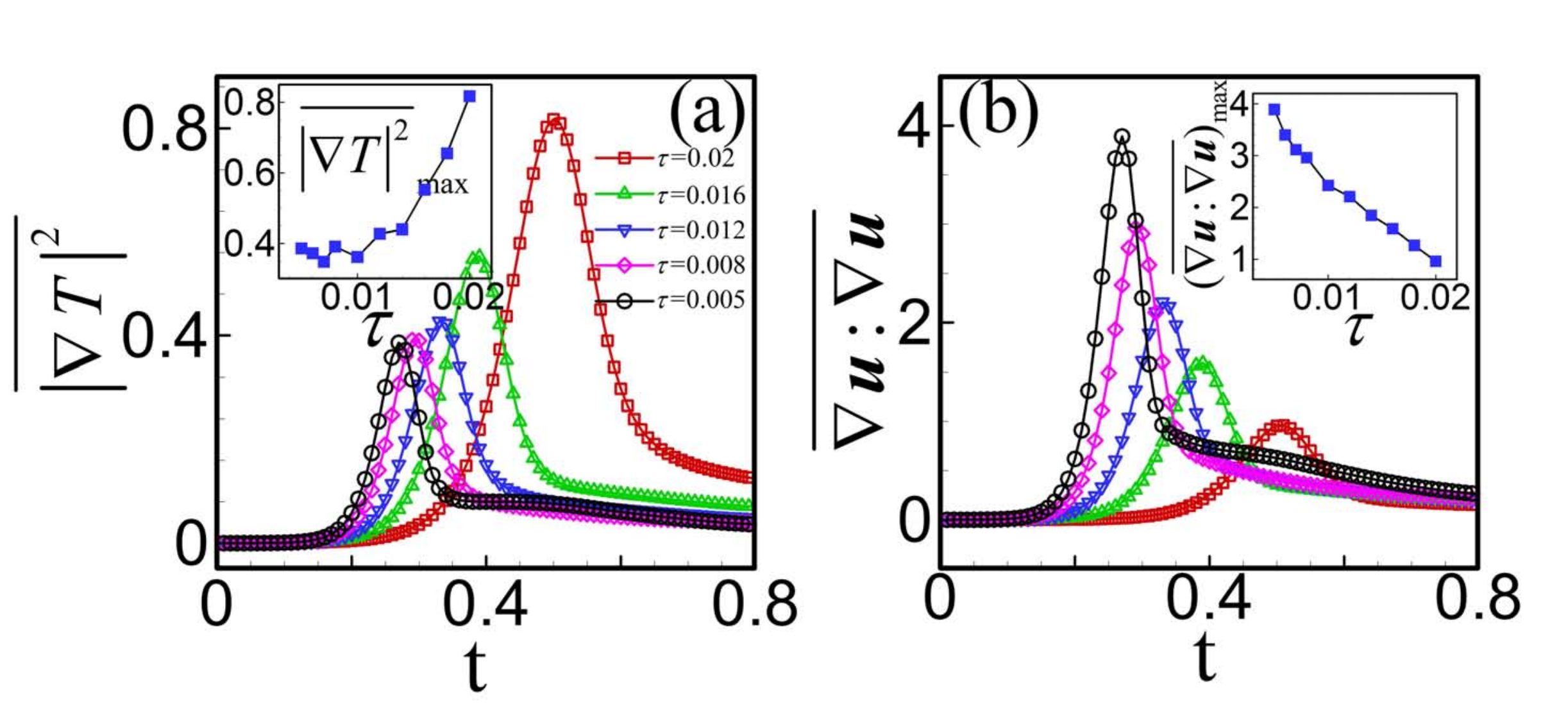}
  \caption{The profiles of the average values of temperature gradient and velocity gradient for different viscosity coefficients, (a) the average value of temperature gradient $\overline {{{\left| {\nabla T} \right|}^2}} $ and (b) the average value of velocity gradient $\overline {\nabla {\bf{u}}:\nabla {\bf{u}}} $.}
  \label{fig5-5}
\end{figure}

 It has been analyzed in the previous section that the characteristics of the entropy production rate of NOEF mainly depends on the temperature gradient while the characteristics of entropy production rate of NOMF are mainly determined by velocity gradient. As the viscosity coefficient increases, the overall motion of the flow field is more remarkable but the mutual motion between different parts of the fluid is much weaker. Consider an extreme case where there is no mutual motion between different regions of the fluid when the viscosity coefficient is infinite. In that case, there is no velocity gradient in the flow field. The profiles of $\overline {\nabla {\bf{u}}:\nabla {\bf{u}}} $ are plotted in Fig. \ref{fig5-5} (b), from which we can see that the profiles for different viscosity coefficients are very similar to those of $\dot S_{NOMF}$ in Fig. \ref{fig5-2} (c). The peak value of $\overline {\nabla {\bf{u}}:\nabla {\bf{u}}} $ as a function of $\tau$ is also plotted in the inset of Fig. \ref{fig5-5} (b), and the peak values decrease with the increase of $\tau$ which is very similar to the decrease of $\dot S^{max}_{NOMF}$ in Fig. \ref{fig5-3} (c). Thus, we can conclude that effects of viscosity is to decrease the velocity gradient in the flow field. Consequently, it decreases the entropy production rate of NOMF.

At the same time, because of the weakening of the mutual motion, the heat convection decreases, thus contributing to the increase of temperature gradient. Figure \ref{fig5-5} (a) shows the profiles of $\overline {{{\left| {\nabla T} \right|}^2}} $ for several different viscosity coefficients. The characteristics of the profiles of $\overline {{{\left| {\nabla T} \right|}^2}} $ are similar to those of $\dot S_{NOEF}$ in Fig. \ref{fig5-2} (b). Besides, the peak value of $\overline {{{\left| {\nabla T} \right|}^2}} $ as a function of $\tau$ is also plotted in the inset of Fig. \ref{fig5-5} (a), which is very similar with the profile of $\dot S^{max}_{NOEF}$ in \ref{fig5-3} (b). Thus, the effect of viscosity is to increase the temperature gradient in the flow field, which leads to a further increase of the entropy production rate of NOEF.

In conclusion, the effect of viscosity is to increase the temperature gradient and decrease the velocity gradient, hence it strengthens the entropy production of NOEF and weakens the entropy production of NOMF.

In addition, the relationship between $\dot S^{max}_{NOEF}$ and $\tau$ can be fitted by a growing exponential function same as that of $t_{SD}$ and the fitting coefficients read $\mathrm{C_1} = 6.09 \times 10^{-4}$, $\mathrm{C_2}=152.91$, and $\mathrm{C_0}=3.7 \times 10^{-3}$. This means that the amplitude of the entropy production rate of NOEF increases exponentially with the coefficient of viscosity. The minimum value of $\dot S^{max}_{NOEF}$ is about $3.7 \times 10^{-3}$, the typical scale of viscosity coefficient is $1/\mathrm{C_2}=0.0065$ which is close to that in Eq. \eqref{Eq-tSD-fitting2}. The amplitude of the exponential growth is about $\mathrm{C_1} = 6.09 \times 10^{-4}$. The relationship between $\dot S^{max}_{NOMF}$ and $\tau$ can be fitted by a linear function with a negative slope, $-0.747$, meaning that the amplitude of the entropy production rate of NOMF decreases linearly with the coefficient of viscosity.

\subsection{Effects of surface tension}
\subsubsection{Thermal phase separation with different coefficients of surface tension}
The surface tension is also an important factor that affects the thermal phase separation.
In this section, the thermal phase separation processes with different coefficients of surface tension $K$ are simulated and investigated.

\begin{figure}[h]
\centering
  \includegraphics[height=6cm]{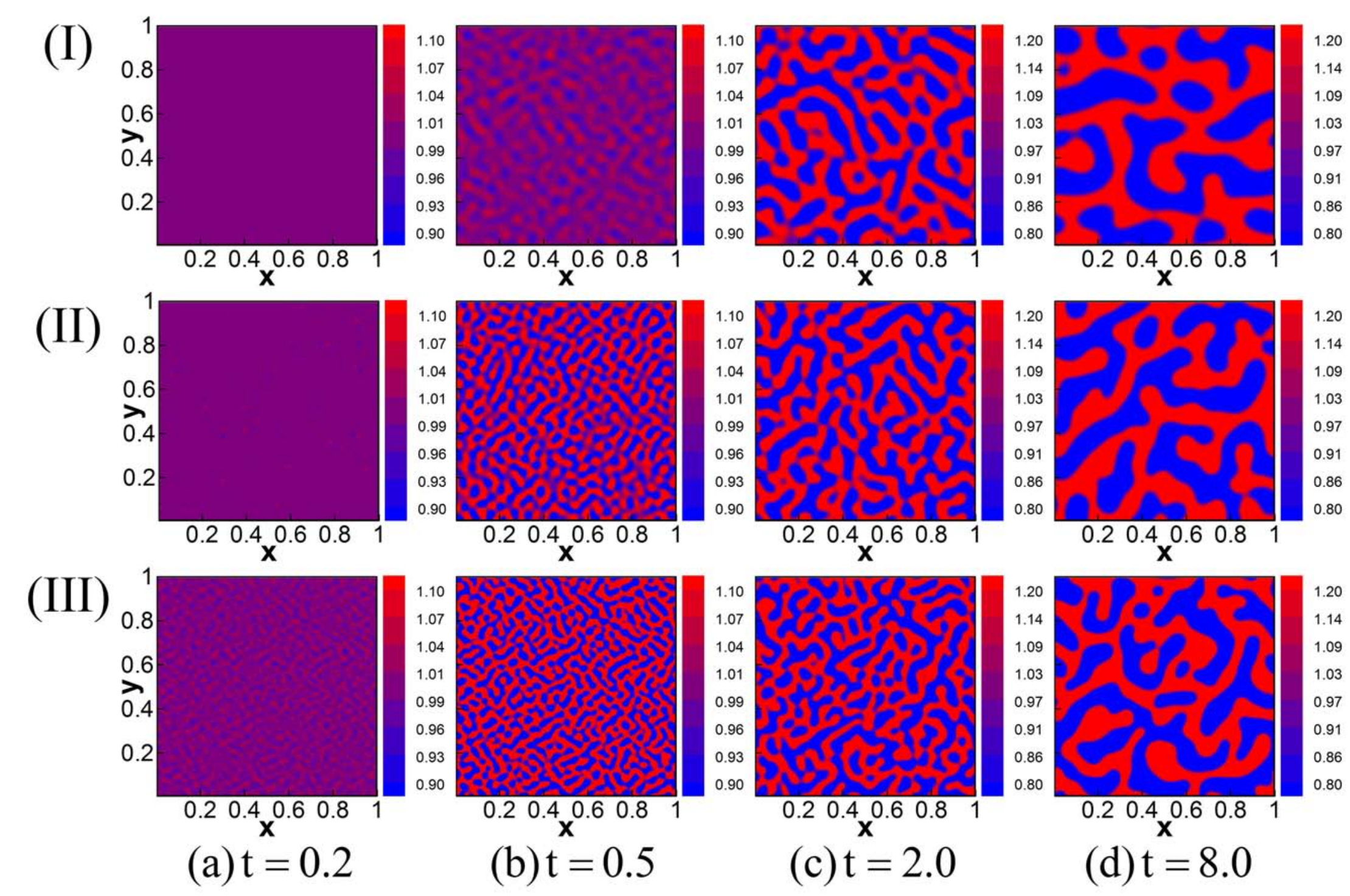}
  \caption{Density contour maps at several times for different values of the surface tension.
The first line corresponds to case (I) $K=2 \times 10^{-5}$, the second line to case (II) $K=1 \times 10^{-5}$, and the third line to case (III) $K=5 \times 10^{-6}$. The columns from left to right are density contour maps at time (a) $t=0.2$, (b) $t=0.5$, (c) $t=2.0$, and (d) $t=8.0$, respectively.}
  \label{fig6-1}
\end{figure}

The simulation conditions are the same as those in Fig. \ref{fig3-1} (II) except for the value of $K$.
Figure \ref{fig6-1} shows the density contour maps with three different $K$ at several time instants.
The three lines from top to bottom correspond to the cases with (I) $K= 2 \times 10^{-5}$, (II) $K= 1 \times 10^{-5}$, and (III) $K= 5 \times 10^{-6}$, respectively.
The four columns from left to right correspond to the snapshots at time (a) $t=0.2$, (b) $t=0.5$, (c) $=2.0$, and (d) $t=8.0$, respectively.
It has been known that the role of surface tension is to promote the merger of the small domains at the DG stage.
From Figs. \ref{fig6-1} (a) and \ref{fig6-1} (b), we can see that the surface tension also plays an important role at the SD stage.
A smaller coefficient of surface tension corresponds to a neater interface and smaller characteristic domain sizes.
Figures \ref{fig6-1} (c) and \ref{fig6-1} (d) belong to the DG stage, in which a smaller surface tension also corresponds to a smaller characteristic domain size.


\begin{figure}[h]
\centering
  \includegraphics[height=7cm]{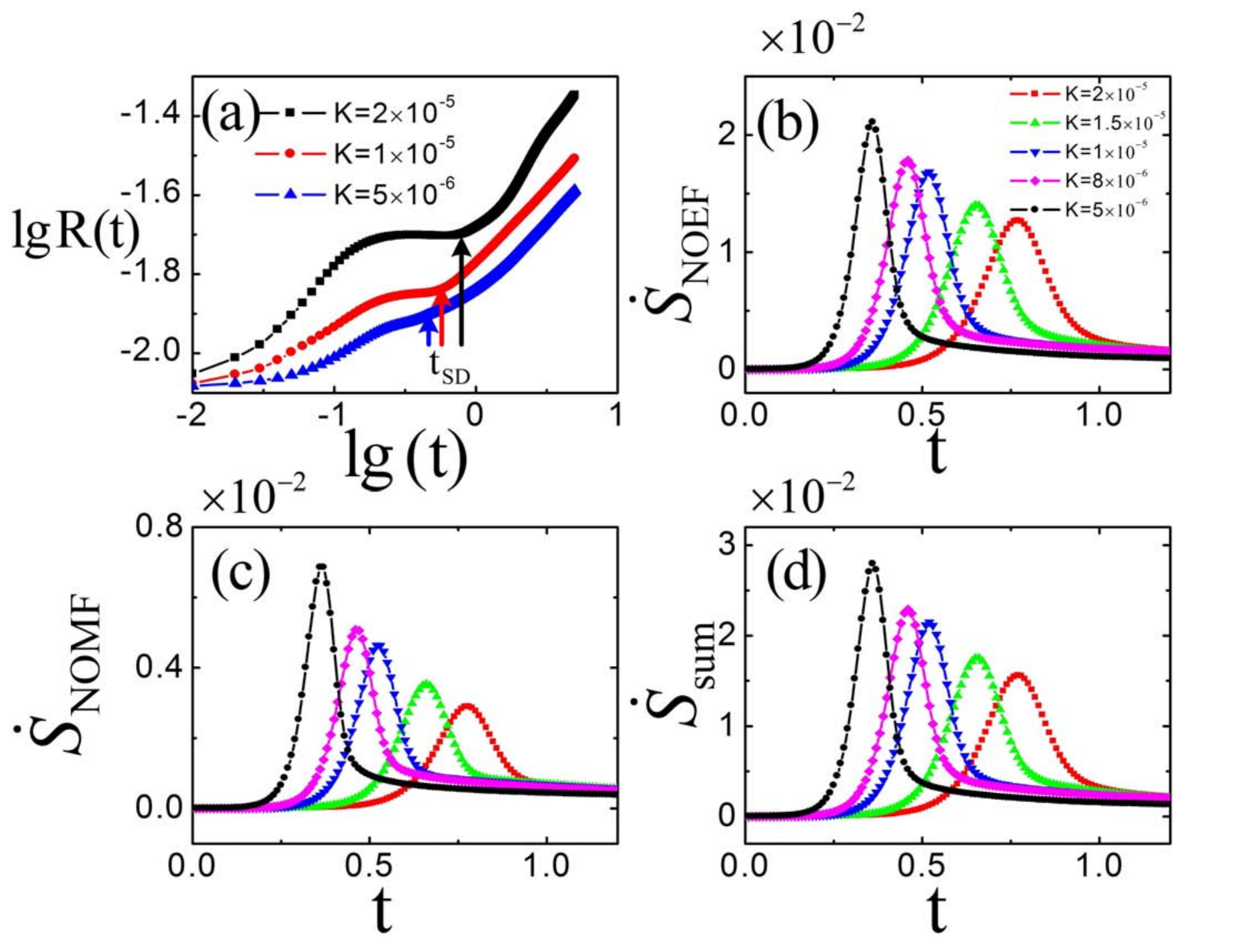}
  \caption{The profiles of characteristic domain sizes and the rates of entropy production for different coefficients of surface tension.
  (a) The characteristic domain sizes $R(t)$ in a log-log scale. (b) The entropy production rate of NOEF $\dot S_{NOEF}$.
  (c) The entropy production rate of NOMF $\dot S_{NOMF}$.
  (d) The total entropy production rate $\dot S_{sum}$.
  The last three subgraphs share the same legend shown in the subgraph (b).}
  \label{fig6-2}
\end{figure}

The evolution of the characteristic domain sizes $R(t)$ over time for several different surface tension coefficients are given
in Fig. \ref{fig6-2} (a). The stronger the surface tension, the larger $R(t)$.
The duration times of the SD stage, $t_{SD}$, are marked by arrows.
The larger values of surface tension associate with longer duration of the SD stage.
Therefore, similar to the viscosity, the role of surface tension is also to inhibit the formation of the phase domains and to prolong the SD stage.
Besides, the surface tension has a significant effect on the plateau of $R(t)$ in the SD stage.
The higher the surface tension, the higher and wider the plateau is.
The plateau nearly vanishes in the case $K=5 \times 10^{-6}$.
The profiles of the entropy production rates, $\dot S_{NOEF}$, $\dot S_{NOMF}$, and $\dot S_{sum}$, for different coefficients of surface tension are shown in Fig. \ref{fig6-2} (b), \ref{fig6-2} (c), and \ref{fig6-2} (d), respectively. The positions of the maximum values of entropy production rates correspond to the points of $t_{SD}$ in Fig. \ref{fig6-2} (a). In addition, all the amplitudes of the entropy production rates including $\dot S_{NOEF}$, $\dot S_{NOMF}$, and $\dot S_{sum}$, increase at decreasing surface tension.

\subsubsection{Effects of surface tension on the SD stage and the entropy production rates}
\begin{figure}[h]
\centering
  \includegraphics[height=3cm]{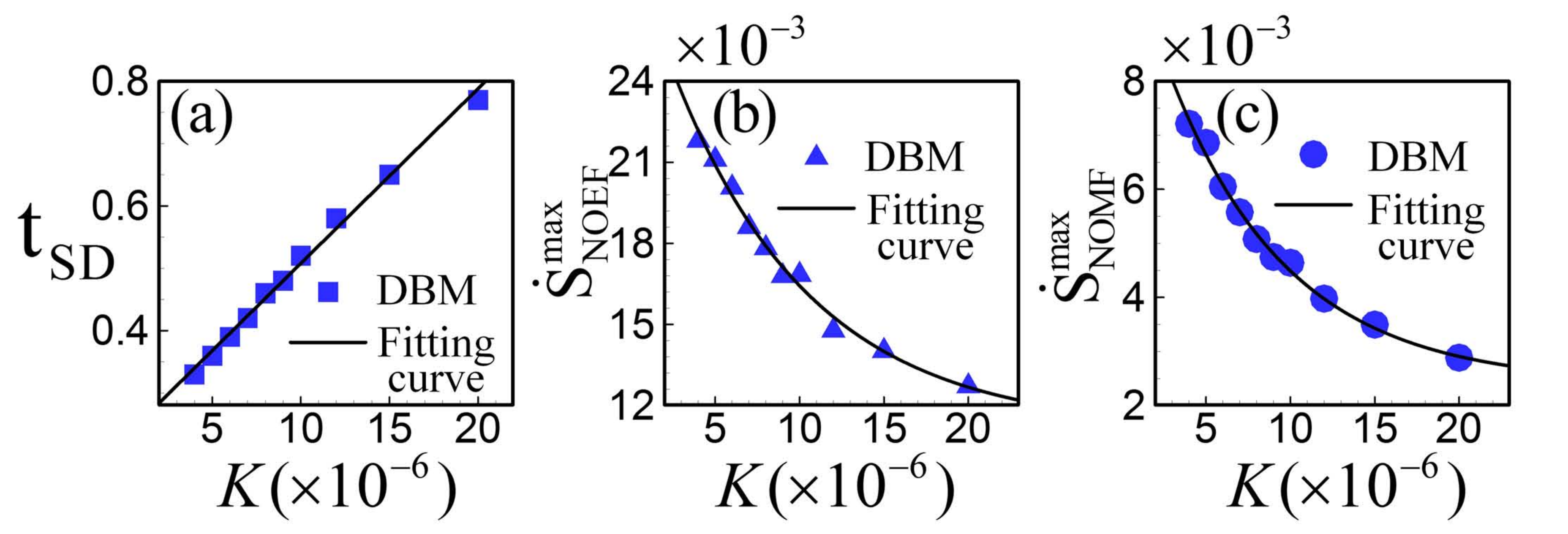}
  \caption{The profiles of duration time of the SD stage, $t_{SD}$, and the amplitudes of entropy production rates as functions of coefficient of surface tension, (a) the duration time of the SD stage $t_{SD}$, (b) the maximum entropy production rate of NOEF $\dot S^{max} _{NOEF}$, and (c) the maximum entropy production rate of NOMF $\dot S^{max} _{NOMF}$. The symbols are results of DBM and the solid lines are fitting curves.}
  \label{fig6-3}
\end{figure}

In order to study the effects of surface tension quantitatively, the profiles of $t_{SD}$, $\dot S^{max}_{NOEF}$, and $\dot S^{max}_{NOMF}$
as a function of $K$ are plotted in Fig. \ref{fig6-3}.
The profile of $t_{SD}$ is given in Fig. \ref{fig6-3} (a) where the symbols represent the results of DBM
and the solid line is fitted by a linear function as
\begin{equation}\label{Eq-surface-tension1}
t_{SD} = \mathrm{C_1} K + \mathrm{C_0},
\end{equation}
where $\mathrm{C_1} = 0.028$ and $\mathrm{C_0} =0.228$. $K$ has a unit of $10^{-6}$ and varies in the range $[4 \times 10^{-6}, 2 \times 10^{-5}]$. $\mathrm{C_0}$ is the value of $t_{SD}$ when the coefficient of surface tension $K$ approximates to zero. As $K$ increases, the duration time of the SD stage increases linearly and the slope is $0.028$. This shows that the dependence on surface tension is much weaker than the corresponding dependence on thermal and momentum conductivities.

The profiles of $\dot S^{max}_{NOEF}$ and $\dot S^{max}_{NOMF}$ are given in Figs. \ref{fig6-3} (b) and \ref{fig6-3} (c), respectively. The symbols represent the results of DBM, where both $\dot S^{max} _{NOEF}$ and $\dot S^{max} _{NOMF}$  are given in units
of $10^{-3}$ and $K$ in units of $10^{-6}$.
It can be seen that both $\dot S^{max}_{NOEF}$ and $\dot S^{max}_{NOMF}$ decrease with increasing surface tension.
The effects of surface tension are to withstand the formation of new interfaces at the SD stage and to facilitate mergers
of small-scale interfaces at the DG stage, so that higher surface tension leads to less interface.
The fewer the phase interfaces, the weaker the effects of temperature gradient and velocity gradient, which are mostly
localized near the interfaces between different phase domains.
As a result, the entropy production rates of NOEF and NOMF both decrease with the increase of surface tension.

\begin{figure}[h]
\centering
  \includegraphics[height=7cm]{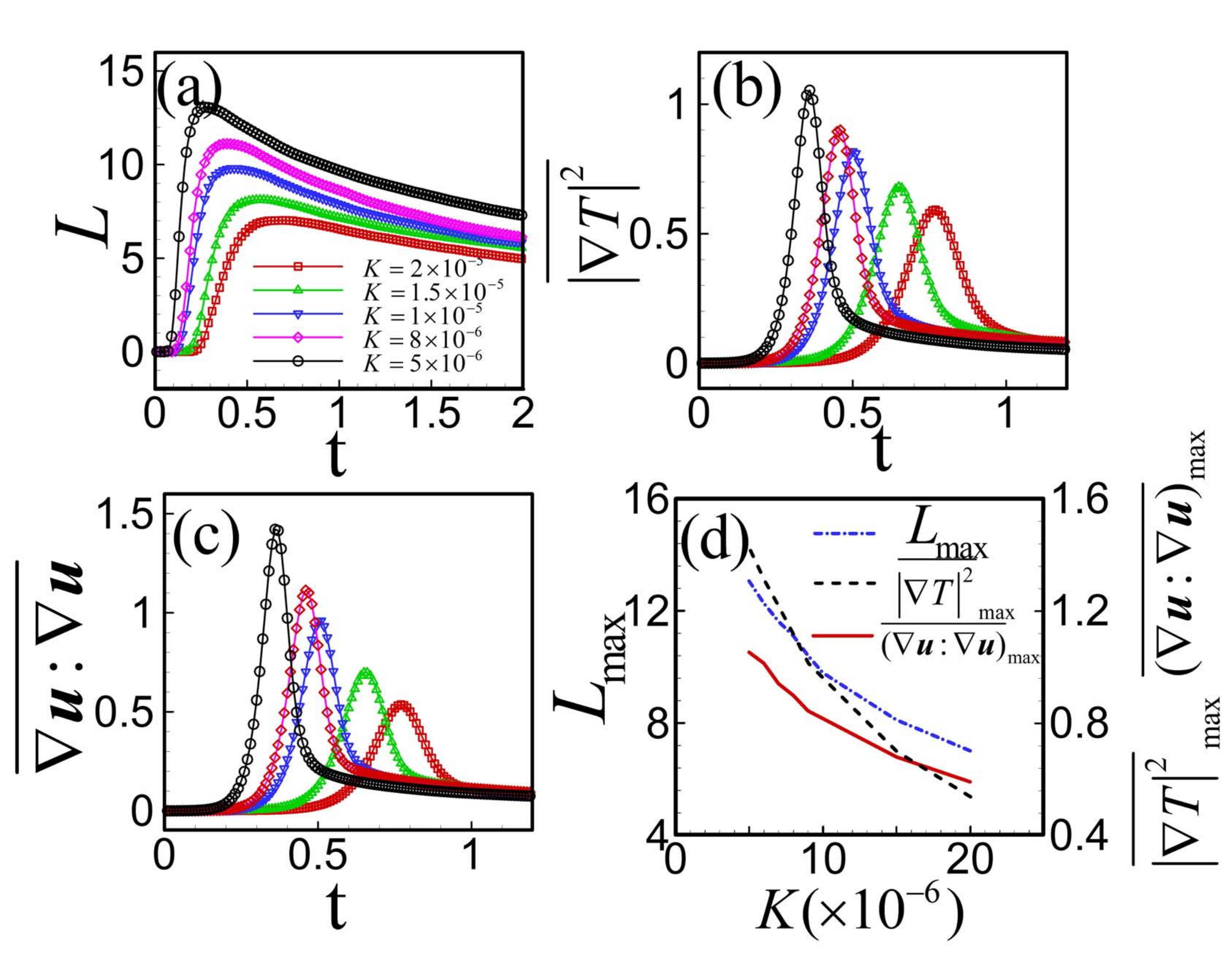}
  \caption{The profiles of morphology boundary length, the temperature gradient, and the velocity gradient gradient for various coefficients of surface tension, (a) Morphology boundary length $L$ as function of time, (b) the temperature gradient $\overline {{{\left| {\nabla T} \right|}^2}}$ as function of time, (c) the velocity gradient $\overline {(\nabla {\bf{u}}:\nabla {\bf{u}})}_{max}$ as function of time, and (d) the peak values $L_{max}$, $\overline {{{\left| {\nabla T} \right|}^2}}_{max}$, and $\overline {(\nabla {\bf{u}}:\nabla {\bf{u}})}_{max}$ as functions of $K$.}
  \label{fig6-5}
\end{figure}

To verify this interpretation, the profiles of the boundary length $L$ for different values of $K$ are plotted in Fig. \ref{fig6-5} (a).
It is apparent that $L$ decreases with increasing $K$. The profiles of $\overline {{{\left| {\nabla T} \right|}^2}} $ and $\overline {\nabla {\bf{u}}:\nabla {\bf{u}}}$ for different values of $K$ are given in Fig. \ref{fig6-5} (b) and Fig. \ref{fig6-5} (c), respectively. These values indicate the average value of the temperature gradient and velocity gradient, respectively.

The characteristics of the profiles of $\overline {{{\left| {\nabla T} \right|}^2}} $ and $\overline {\nabla {\bf{u}}:\nabla {\bf{u}}}$ are very similar to those of $\dot S_{NOEF}$ and $\dot S_{NOMF}$ in Figs. \ref{fig6-2} (b) and \ref{fig6-2} (c), respectively.
The peak values of $L$, $\overline {{{\left| {\nabla T} \right|}^2}} $, and $\overline {\nabla {\bf{u}}:\nabla {\bf{u}}}$ as functions of $K$ are plotted in Fig. \ref{fig6-5} (d), from which we can see that, along with the decrease of $L$, both the temperature gradient and the velocity
gradient decrease with increasing $K$.
The profiles of $\overline {{{\left| {\nabla T} \right|}^2}} _{max}$ and $\overline {(\nabla {\bf{u}}:\nabla {\bf{u}})}_{max}$ are very similar
to the profiles of $\dot S^{max}_{NOEF}$ and $\dot S^{max}_{NOMF}$ in Figs. \ref{fig6-3} (b) and \ref{fig6-3} (c), respectively.
This shows that surface tension decreases the entropy production rates of NOEF and NOMF, by decreasing the length of the interface
between different domains. In addition, the relationship between the amplitudes of both entropy production rates, $\dot S^{max}_{NOEF}$ and $\dot S^{max}_{NOMF}$,  and the coefficient of surface tension can be fitted by a decaying exponential function,
  \begin{equation}\label{Eq-SNOEF-fitting3}
\dot S_{NOEF/NOMF}=\mathrm{C_1} \exp (-\mathrm{C_2} K)+\mathrm{C_0},
\end{equation}
where $\mathrm{C_1}=18.0$, $\mathrm{C_2}=0.122$, and $\mathrm{C_0}=11.1$ for $\dot S_{NOEF}$, and $\mathrm{C_1}=8.60$, $\mathrm{C_2}=0.141$, and $\mathrm{C_0}=13.5$ for $\dot S_{NOMF}$. Combined with the fitting results, we observe that both amplitudes
of $\dot S_{NOEF}$ and $\dot S_{NOMF}$ decrease exponentially with increasing surface tension. When the coefficient of surface tension tends to infinity, both $\dot S_{NOEF}$ and $\dot S_{NOMF}$ attain their minimum, $\dot S_{NOEF}=11.1$ and $\dot S_{NOMF}=13.5$. Oppositely, as the coefficient of surface tension tends to zero,  both $\dot S_{NOEF}$ and $\dot S_{NOMF}$ attain their maximum, $\dot S_{NOEF}=29.1$ and $\dot S_{NOMF}=22.1$. The typical scales
of coefficient of surface tension for $\dot S_{NOEF}$ and $\dot S_{NOMF}$  are $1/\mathrm{C_2}$, i.e., $8.20$ and $7.09$, respectively, which are close to each other. However, from the amplitude of the exponential decay, $\mathrm{C_1}$, we find that the effect of surface tension on entropy production of NOEF is stronger than that of NOMF.

\subsection{Cooperation and competition between NOEF and NOMF for entropy production}

In this section, we study the cooperation and competition between the two mechanisms,
NOEF and NOMF, in the entropy production rate under various heat conductivities, viscosity coefficients, and coefficients of surface tension.

\begin{figure}[h]
\centering
  \includegraphics[height=2.8cm]{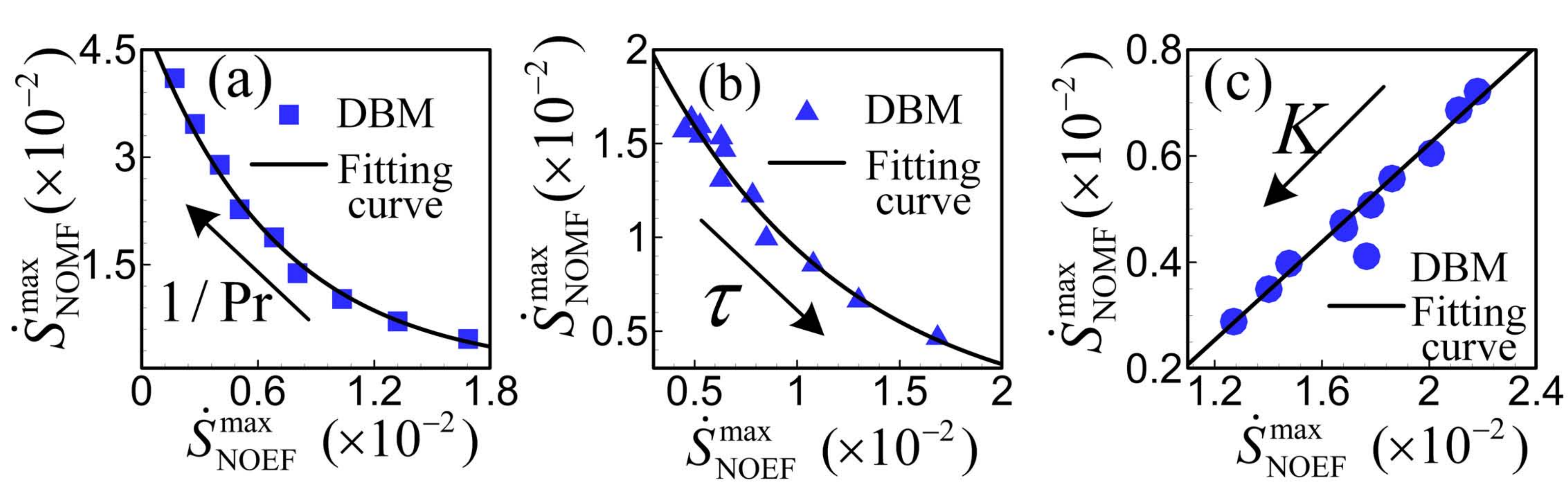}
  \caption{The maximum entropy production rate $\dot S^{max}_{NOMF}$ versus $\dot S^{max}_{NOEF}$ under various (a) heat conductivities, (b) viscosity coefficients, and (c) coefficients of surface tension. The arrows in subfigures (a), (b), and (c) indicate an increase in heat conductivity, viscosity coefficient, and coefficient of surface tension, respectively. The symbols are DBM results and the solid lines are fitting curves.}
  \label{Phase-fig1}
\end{figure}

The profile of $\dot S^{max}_{NOMF}$ versus $\dot S^{max}_{NOEF}$ under the change of heat conductivity is plotted in Fig. \ref{Phase-fig1} (a).
The arrow points to the direction along which heat conductivity (represented by $1/Pr$) increases.

The symbols are DBM results, from which we can see that there is a competition between $\dot S^{max}_{NOMF}$ and $\dot S^{max}_{NOEF}$ as the heat conductivity changes. As heat conductivity increases, $\dot S^{max}_{NOMF}$ increases while $\dot S^{max}_{NOEF}$ decreases.
Conversely, the former decreases while the latter increases.
The relationship between $\dot S^{max}_{NOMF}$ and $\dot S^{max}_{NOEF}$ can be fitted by
\begin{equation}\label{Eq-Phase1}
\dot S^{max} _{NOMF} = \exp (- \mathrm{C_1} \dot S^{max} _{NOEF} + \mathrm{C_0}).
\end{equation}
where $\mathrm{C_1}=1.47$ and $\mathrm{C_0}=1.61$.

From the fitting curve, we learn that the increase (decrease) of the $S^{max}_{NOMF}$ is exponentially related to the decrease (increase) of the $S^{max}_{NOEF}$.
From Fig. \ref{Phase-fig1} (b), we note a similar relationship between $\dot S^{max}_{NOMF}$ and $\dot S^{max}_{NOEF}$, under the change of viscosity coefficient. The arrow in the figure points to the direction along which the coefficient of viscosity increases.
The symbols are DBM results and can be fitted by a similar expression as shown in Eq. \eqref{Eq-Phase1} and the coefficients read $\mathrm{C_1}=1.06$ and $\mathrm{C_0}=0.997$.
As the coefficient of viscosity increases, $\dot S^{max}_{NOMF}$ decreases, while $\dot S^{max}_{NOEF}$ increases.
The decrease of the former is exponentially related to the increase of the latter.
The competitive relationship between $\dot S^{max}_{NOMF}$ and $\dot S^{max}_{NOEF}$ is similar to that
in Fig. \ref{Phase-fig1} (a), except that $\dot S^{max}_{NOMF}$ and $\dot S^{max}_{NOEF}$ change in opposite directions.

In fact, both the decrease of heat conductivity and the increase of coefficient of viscosity are equivalent to an increase of Prandtl number.
Thus, combining Figs. \ref{Phase-fig1} (a) and \ref{Phase-fig1} (b), we find that $\dot S^{max}_{NOMF}$ decreases and $\dot S^{max}_{NOEF}$ increases with the increase of Prandtl number.

Generally, the increase of Prandtl number indicates the relative strengthening of the viscosity and the weakening of the heat conduction.
However, it is interesting to note that an increasing Prandtl number corresponds to the decrease of $\dot S^{max}_{NOMF}$ and the increase of $\dot S^{max}_{NOEF}$. Based on the previous analysis, we conclude that the entropy production rates $\dot S_{NOEF}$ and $\dot S_{NOMF}$ are mainly determined by the temperature and velocity gradients, respectively.
With the increase of viscosity coefficient or the decrease of heat conductivity, the velocity gradient decreases, while
the temperature gradient increases. Consequently, the increase of Prandtl number
contributes to the decrease of $\dot S^{max}_{NOMF}$ and to the increase of $\dot S^{max}_{NOEF}$.

Figure \ref{Phase-fig1} (c) shows the profile of $\dot S^{max}_{NOMF}$ versus $\dot S^{max}_{NOEF}$ for various coefficients of surface tension.
The arrow in the figure points to the direction along which coefficient of surface tension increases. The symbols are DBM results, from which we can see that $\dot S^{max}_{NOMF}$ and $\dot S^{max}_{NOEF}$ drop synchronously with the increase of coefficient of surface tension. The relationship between $\dot S^{max}_{NOMF}$ and $\dot S^{max}_{NOEF}$ under different values of $K$ can be fitted by a linear function, with positive slope, which reads
 \begin{equation}\label{Eq-Phase3}
\dot S^{max} _{NOMF} = \mathrm{C_1} \dot S^{max} _{NOEF} + \mathrm{C_0},
\end{equation}
The slope is $\mathrm{C_1}=0.466$.
From the fitting results, we observe that, as the coefficient of surface tension increases,  both $\dot S^{max}_{NOMF}$ and $\dot S^{max}_{NOEF}$ decrease and the reduction of $\dot S^{max}_{NOMF}$ is proportional to that of $\dot S^{max}_{NOEF}$. Besides, since the slope is less than $1$, the reduction of $\dot S^{max}_{NOMF}$ is less than that of $\dot S^{max}_{NOEF}$. Thus, we can conclude that the effect of surface tension is to decrease the entropy production rate, and it mainly leads to the cooperation, instead of competition between $\dot S^{max}_{NOMF}$ and $\dot S^{max}_{NOEF}$. The reason is mainly attributed to the shrinking of the phase interfaces. As surface tension increases, the length of the phase interface decreases. Since the temperature and velocity gradients
are localised near the interfaces between different phase domains, they decrease together with the decrease of the length of the phase interface.

\section{Conclusions}

Based on the multiphase flow DBM, we have investigated the entropy production of thermal phase separation and
established the relationship between entropy production rate and the
non-organised energy fluxes (NOEF) and the non-organised moment fluxes (NOEF).

A new physical criterion to separate the two stages of phase separation is presented.
It is found that the entropy production rate increases with time at the spinodal decomposition stage and decreases
with time in the domain growth stage, the maximum of the entropy production rate corresponding to the critical time $t_{SD}$ marking the transition between the two stages.

Then, the effects of heat conduction, viscosity, and surface tension on the entropy production are examined.

Within the range of simulation parameters explored in this work, it is found that the amplitude of the entropy production rate
of NOEF decreases at increasing heat conductivity, increases at increasing viscosity coefficient and decreases at increasing surface tension. The amplitude of the entropy production rate of NOMF increases with increasing  heat conductivity, decreases with increasing viscosity coefficient and decreases with increasing of surface tension.

This can be explained from inspection of the temperature and velocity gradients in the flow field, which are the dominant
factors affecting the entropy production rates of NOEF and NOMF, respectively.
With the increase of heat conductivity or the decrease of the viscosity coefficient, the temperature gradient decreases, while
the velocity gradient increases, which results in the decrease of the NOEF entropy production rate and the increase the NOMF entropy production rate.
As the coefficient of surface tension increases, the length of the interface between different phase domains decreases, which
leads to the decrease of both temperature gradient and velocity gradient, since they are most located at the interface.
Consequently, the increase of the coefficient of surface tension leads to the decrease of both NOEF and NOMF entropy production rates.

The amplitude of sum entropy production rate is a combination of NOEF and NOMF contributions.
In addition, it is found that the competition between the NOMF and NOEF entropy production is much more substantial with respect to changes of the heat conductivity or viscosity coefficient.
This means that the increase (decrease) of the entropy production rate due to NOMF corresponds to the decrease (increase)
of the entropy production rate of NOEF. However, both NOMF and NOEF  entropy production rate change
synchronously with surface tension, i.e. they cooperate instead of competing.

\section*{Acknowledgements}
YZ, AX, and GZ acknowledge the support of National Natural Science Foundation of China (under grant no. 11772064), CAEP Foundation under Grant No. CX2019033, Science Challenge Project (Grant No. JCKY2016212A501), the opening project of State Key Laboratory of Explosion Science and Technology (Beijing Institute of Technology) under Grant No. KFJJ19-01 M.
ZC acknowledges the support of National Natural Science Foundation of China (under grant no. 11502117).
YG acknowledges the support of National Natural Science Foundation of China (under grant no. 11875001), Natural Science Foundation of Hebei Province (under grants no. A2017409014) and Natural Science Foundations of Hebei Educational Commission (under grant no. ZD2017001).
S.S. wishes to acknowledge funding from the European Research Council under the European Union's Horizon 2020 Framework Programme (No. FP/2014-2020)/ERC Grant Agreement No. 739964 (COPMAT).


\bibliography{rsc} 
\bibliographystyle{rsc} 

\end{document}